\documentstyle[12pt,epsfig,epsf]{article}
\def\Hpm{ {H^{\pm}}}
\def\h0{{h^0}}
\def\H0{{H^0}}

\begin{document}
\newcommand{\nn}{\noindent}
\renewcommand{\thefootnote}{\fnsymbol{footnote}}
\begin{flushright}
LPM/99-61\\
UFR-HEP/99/01 \\
KA-TP-21-1999 \\
November 1999 \\
\end{flushright}
\vspace*{.4cm}
\begin{center}
{\large{ {\bf Associated H$^{-}$\,W$^{+}$ Production}}}
{\large{ {\bf in High Energy  $e^+e^-$ Collisions}}}
\vspace{.4cm}

{{\sc A. Arhrib}$^{\mbox{1,2}}$,
{{\sc M. Capdequi Peyran\`ere}$^{\mbox{3}}$}, 
{{\sc W. Hollik}$^{\mbox{4}}$} and {{\sc G. Moultaka}$^{\mbox{3}}$}}

\vspace{.6cm}
1: D\'epartement de Math\'ematiques, Facult\'e des Sciences et Techniques\\
B.P 416, Tanger, Morocco

\vspace{.6cm}
2: UFR--High Energy Physics, Physics Department, Faculty of Sciences\\
PO Box 1014, Rabat--Morocco

\vspace{.6cm}
3: Laboratoire de Physique Math\'ematique et Th\'eorique, CNRS-UMR 5825\\
Universit\'e Montpellier II, F--34095 Montpellier Cedex 5, France

\vspace{.6cm}
4: Institut f\"ur Theoretische Physik,Universit\"at Karlsruhe\\
D-76128 Karlsruhe, Germany \\

\end{center}

\setcounter{footnote}{4} 
\vspace{1.2cm}

\begin{abstract}
\nn We study the  associated production of charged Higgs bosons with  
$W$ gauge bosons in high energy $e^+ e^-$ collisions at 
 the one loop level.  
We present the analytical results and give a
detailed discussion for the
total cross section predicted 
in the context of a general
Two Higgs Doublet Model (THDM).  
\end{abstract}

\newpage

\pagestyle{plain}
\renewcommand{\thefootnote}{\arabic{footnote} }
\setcounter{footnote}{0}

\section*{1.~Introduction}
The discovery of the standard Higgs boson is one of the major goals of 
the present  and future
searches in particle physics. Direct or indirect 
results give stringent lower and upper bounds on its mass 
\cite{Hmasses1,Hmasses}.
 Moreover, the problematic scalar sector of the Standard Model (SM) can be 
enlarged and some simple extensions such as the minimal 
Two Higgs Doublet Model (THDM) versions \cite{gun} are intensively studied. 
The two most popular versions of the THDM (type I and II), differ in some 
Higgs couplings to fermions, but in both types and after
electroweak symmetry breaking \cite{ewsb}, 
the Higgs spectrum is the same. From 
the 8 degrees of freedom initially present in the 
2 Higgs doublets, 3 correspond to masses of the longitudinal gauge bosons
and we are left with 5 degrees of freedom which manifest themselves as 5 
physical Higgs particles (2 charged Higgs $H^\pm$, 2 CP-even $H$, $h$
and one CP--odd $A$). 
The charged Higgs $H^\pm$, 
because of its electrical charge, is noticeably different from the 
other SM or THDM Higgs particles. In other words the discovery 
of a charged scalar Higgs boson would attest definitively that the 
Standard Model is overcome. On the contrary a neutral Higgs 
experimental evidence has to be refined to go 
beyond the SM\cite{Hmasses,Hneutre}.

Furthermore, one can construct models with an even larger scalar sector than 
in the THDM, one of the most popular being the Higgs Triplet Model (HTM)
\cite{triplet}. A noteworthy difference between THDM and HTM is that 
the HTM contains a tree level $ZW^\pm H^\mp$ coupling  while in 
the THDM this coupling is only generated at one loop level \cite{gun, micapey}. In models of the HTM type, the physical predictions 
are sometimes very different from typical ones in the SM and or the
THDM.  In our study, we will only consider models with two 
Higgs doublets.

The study of THDM is also very interesting insofar as it provides a general 
framework to test the minimal suspersymmetric extension of the SM 
(MSSM) \cite{MSSM}. Of course the particle spectrum of the MSSM doubles that
 of the THDM, but the scalar Higgs sectors, 
responsible for the electroweak symmetry breaking, are analogous, 
up to some mass constraints and couplings in the MSSM. 
In any case, the study of the various production mechanisms 
of charged Higgs bosons would give information about new physics beyond the SM.

In future colliders or linear accelerators devoted to the electron - 
positron annihilations, 
the simplest way to get a charged Higgs is through  
$H^\pm$  pair production. 
Such studies have been 
already undertaken at tree-level \cite{komamiya}
and one-loop orders  \cite{ACM, AM} and shown that
$e^+e^-$ machines will offer a clean environment and in that sense 
a higher mass reach, especially if the 1--2 TeV options are available.
We mention also that charged Higgs bosons pair production  
through laser back--scattered $\gamma\gamma$ 
collisions has been studied in the 
literature \cite{pp} and found to be prominent to discover the
charged Higgs boson.

Note also that charged Higgs bosons can be produced in hadronic machines:

($i$) single charged  Higgs production via $gb \to tH^-$,
$gg\to t\bar{b}H^-$, $qb\to q' bH^-$ \cite{single}

($ii$) single charged Higgs production in association with 
$W$ gauge boson via
$ gg \to W^\pm H^\mp $ or $b\bar{b} \to  W^\pm H^\mp $ \cite{wh}

($iii$) $H^\pm$ pair production through $q \bar{q}$ 
annihilation would be feasible 
only if the charged Higgs decay in top-bottom 
is kinematically forbidden \cite{hadronic},
otherwise one would have to look at bosonic decay modes 
(ex. $H^\pm \rightarrow h^0 W^\pm$, 
$H^\pm \rightarrow A^0 W^\pm$, etc...).

The aim of this paper is the study of the  production of the
charged Higgs boson in association with a $W$ gauge boson 
through electron positron annihilation. 
This process is kinematically better suited
than the $H^\pm$ pair production when 
the Higgs mass is larger than the 
W mass. Although it is a rare process
in the THDM, 
loop or/and threshold effects 
can give a substantial enhancement.
Moreover, once worked out, any experimental deviation from the results 
within such a model should bring some fruitful information on the new 
physics.   

The paper is organized as follows. In section II, we review all the 
charged Higgs boson interactions (with gauge bosons, scalar 
bosons and fermions), section III contains the notations and 
conventions while section IV is devoted to the on--shell 
renormalization scheme 
we will use. In section V we present our numerical results, 
and section VI contains our conclusions.

\renewcommand{\theequation}{2.\arabic{equation}}
\setcounter{equation}{0}
\section*{2. Charged Higgs  boson interactions }
\subsection*{2.1 Charged Higgs  boson interactions with scalar bosons}
 In this section we will follow the notation of ref \cite{GH}. 
Recalling 
the most general (dimension 4) $SU(2)_{weak} \times U(1)_Y$ 
gauge-invariant and CP-invariant scalar potential for the THDM
\begin{eqnarray}
& & V(\Phi_{1},\Phi_{2})=\lambda_{1} (\Phi_{1}^+\Phi_{1}-v_{1}^2)^2
+\lambda_{2} (\Phi_{2}^+\Phi_{2}-v_{2}^2)^2+
                    \lambda_{3}((\Phi_{1}^+\Phi_{1}-v_{1}^2)+(\Phi_{2}^+
\Phi_{2}-v_{2}^2))^2                 \nonumber\\ [0.2cm]
                    & &+\lambda_{4}((\Phi_{1}^+\Phi_{1})(\Phi_{2}^+
\Phi_{2})-(\Phi_{1}^+\Phi_{2})(\Phi_{2}^+\Phi_{1}))+
                    \lambda_{5} (Re(\Phi^+_{1}\Phi_{2})
-v_{1}v_{2})^2+\nonumber\\ [0.2cm]
                    & & \lambda_{6}
[Im(\Phi^+_{1}\Phi_{2})]^2 +\lambda_{7} \, ,
\label{higgspot}
\end{eqnarray}
  
we have the doublet fields $\Phi_1$ and $\Phi_2$ with 
weak hypercharge Y=1,  
the corresponding vacuum expectation values $v_1$ and $v_2$,
and the coefficients $\lambda_i$
as real-valued parameters.
We  will assume 
the arbitrary additive constant $\lambda_{7}$ to be vanishing.
 Via the Higgs mechanism, the W and Z gauge 
bosons acquire masses  
given by  $m_W^2=\frac{1}{2}g^2 v^2$ and $m_Z^2= \frac{1}{2}(g^2 +g'^2) v^2$,
where $g$ and $g'$ are the $SU(2)_{weak}$ and $U(1)_Y$ gauge couplings and
$ v^2= v_1^2 + v_2^2$. The combination $v_1^2 + v_2^2$ 
is thus fixed by the electroweak 
scale through $v_1^2 + v_2^2=(2\sqrt{2} G_F)^{-1}$, 
and we are left with 7 free parameters in eq.(\ref{higgspot}), 
 namely $\lambda_i$ and $\tan\beta=v_2/v_1$. 
 As shown in \cite{AM}, using straight-forward algebra, one can 
relate the coefficients $\lambda_i$  of the scalar potential
to the masses of the 
physical Higgs bosons $h,\,H,\,A,\,H^{\pm}$ \cite{GH} 
in the following way:   
\begin{eqnarray}
& & \lambda_1= \frac{g^2}{16 \cos^2\beta m^2_W} [m^2_H+m^2_h +
(m^2_H - m^2_h)\frac{\cos(2\alpha +\beta)}{\cos\beta}]
+ \lambda_3(-1 + \tan^2\beta) \label{lambda1} \\
&&\lambda_2=\frac{g^2}{16 \sin^2\beta m^2_W} [m^2_H+m^2_h +
(m^2_h - m^2_H)\frac{\sin(2\alpha +\beta)}{\sin\beta}]
+ \lambda_3(-1 + \cot^2\beta) \label{lambda2}
\end{eqnarray}
\begin{eqnarray}
& & \lambda_4=\frac{g^2m_{H^\pm}^2}{2 m^2_W}  \quad , \quad
\lambda_5=\frac{g^2}{2 m^2_W} \frac{\sin 2\alpha}{ \sin 2 \beta} 
(m_H^2-m_h^2)\ -\ 4 \lambda_3
\quad , \quad \lambda_6= \frac{g^2m_{A}^2}{2 m^2_W}  
 \label{lambda5} \\ \nonumber
\end{eqnarray}
Consequently one can choose as  
free parameters of the Higgs sector: 
 the four Higgs boson masses 
($m_{H^\pm}$, $m_H$, $m_h$, $m_A$), 
$\alpha$, $\tan\beta$ and $\lambda_3$.

The trilinear self couplings required for our study are listed in 
the appendix A, Eqs.~(\ref{A6})--(\ref{H2GPGM}). 
The trilinear vertices $H^0H^+H^-$ and 
$h^0H^+H^-$ depend, besides on the Higgs boson masses, 
on $\lambda_3$ and $\tan\beta$. From
Appendix A, eqs.(\ref{A6}, \ref{A7}),
it is obvious that one would get into conflict with the requirements
from perturbative 
unitarity when 
$\tan\beta$ and/or $\lambda_3$ are large.
In our analysis
we will take into account the following constraints
when the masses and the coupling parameters are varied:
\begin{itemize}
\item  Unitarity constraints 
will be respected in a simplified way, following \cite{unitarity}, 
by imposing the condition 
\begin{equation}
 \vert HHH\vert < \frac{3 g}{2m_W} (1 \mbox{TeV})^2 \label{UNITARITY}
\end{equation}
on each trilinear scalar vertex $HHH$ entering the process at one loop.  
\item  Lower bounds on $\tan\beta$ in the general THDM
 have been obtained from the experimental limits on the processes 
 $e^+e^-\to Z^*\to h^0\gamma$ and/or $e^+e^-\to A^0\gamma$ 
\cite{kraw}.
For light $h$ masses, these bounds can be rather low
\cite{kraw}. 
For our study we will restrict the discussion to values 
$\tan\beta>0.5$.
\item 
 The extra contribution $\delta\rho$ to the $\rho$ parameter 
\cite{dhk}  should not exceed the 
current limits from precision measurements \cite{el}:
$$-0.0017 \leq \delta\rho \leq 0.0027 $$
\end{itemize}

\subsection*{2.2 Charged  Higgs boson interactions with fermions}
In the  two-Higgs-doublet extension of the Standard Model 
there are different ways to couple the Higgs fields to the fermions. 
 Conventionally they are classified in terms of the following
categories, labeled as type I and type II models:
\begin{itemize}
\item[i.] {\sf Model type I:}  All quarks and leptons
couple exclusively to the second Higgs doublet $\Phi_2$, with the 
coupling structure copied from the Standard Model. 
$\Phi_2$ gives mass to both up- and down-type quarks,
invoking 
the charge-conjugate of $\Phi_2$ for the up-type quarks. 
Since supersymmetry forbids the appearance of the complex conjugate,  
type I models cannot be realized within the MSSM.   
\item[ii.] {\sf Model type II:} To avoid the problem of
 flavor changing neutral currents (FCNC) \cite{glashow-weinberg}, 
one assumes 
that $\Phi_1$ couples only to down-type quarks and charged leptons
and $\Phi_2$ to up-type quarks (and eventually to neutrinos).
The type II model is the pattern found in the MSSM. 
\end{itemize}
 The general structure of 
the charged-Higgs boson interaction with
a doublet of up- and down-type fermions is given by the vertex 
\begin{equation}
[H^- u\bar d] = \frac{ig V_{ud}}{ \sqrt{2} m_W } \{ Y_{ud}^L
\frac{(1-\gamma_5)}{2} + Y_{ud}^R \frac{(1+\gamma_5)}{2}\} \, .
\label{27}
\end{equation}
 In the specific models mentioned above, the Yukawa
couplings have the form
\begin{eqnarray} 
& & Y_{ud}^L=\frac{ m_u}{\tan\beta} \qquad
\mbox{ and} \qquad 
 Y_{ud}^R= -\frac{ m_d }{ \tan\beta}\qquad \mbox{for  model I} \, , 
\nonumber \\              
 & & Y_{ud}^L=\frac{ m_u}{ \tan\beta} \qquad \mbox{ and} 
\qquad
Y_{ud}^R=   m_d \tan\beta\qquad \mbox{for  model II} \, . 
\end{eqnarray}
$V_{ud}$ is the CKM matrix element, which we will 
approximate by unity.
Models I and II lead to similar results as
far as the effects originating from the
top--bottom loops 
for $\tan\beta$ close to 1 are considered,
owing essentially to the form 
of the Yukawa couplings $Y^R$ and $Y^L$. 
 
\subsection*{2.3 Charged  Higgs-boson interactions with gauge bosons }
 The interactions 
between the charged Higgs bosons of a two-doublet model and 
the gauge bosons are completely dictated by 
local gauge invariance and thus independent of the assumption  
whether the model is
supersymmetric or not. These 
interactions follow from the kinetic term, involving the
covariant derivative, in the Higgs 
Lagrangian 
\begin{eqnarray}
&  & \sum_i (D_{\mu}\Phi_i)^+(D_{\mu}\Phi_i)  =  \nonumber   \\  
&  & \sum_i [(\partial_\mu +ig \vec{T_a} 
\vec{W_{\mu}^a}  +ig'\frac{Y_{\Phi_i}}{2}B_\mu) \Phi_i]^+[(\partial_\mu +ig 
\vec{T_a} \vec{W_{\mu}^a} +ig'\frac{Y_{\Phi_i}}{2}B_\mu )\Phi_i] \, ,
\label{covder} 
\end{eqnarray}
where $\vec{T_a}$ is the isospin operator and $Y_{\Phi_i}$ 
the hypercharge of the Higgs fields;  
${W^a}_\mu$ denote the $SU(2)_L$ gauge fields,
$B_\mu$ the $U(1)_Y$ gauge field, and $g$ (resp. $g'$) the associated coupling
constants.
The individual couplings in the various vertices, derived from 
eq.~(\ref{covder}), are listed in Appendix C.
The $\gamma H^\mp W^\pm$ and the $Z H^\mp W^\pm$  
vertices vanish at the tree level. 
Therefore, tree-level contributions to $e^+e^- \to W^\pm H^\mp$ 
come only from neutrino exchange in the $t$--channel  and from 
CP-even Higgs mediated s-channel. 
All these contributions are strongly
suppressed by the small electron mass. 
Hence, the process under study is essentially a loop-mediated process. 

\renewcommand{\theequation}{3.\arabic{equation}}
\setcounter{equation}{0}

\section*{3. Notations and conventions}
We will use the following notations and conventions.
The momenta of the incoming electron and positron, outgoing
gauge  boson $W^+$ and outgoing Higgs  boson 
$H^-$ are denoted by $p_1$, $p_2$, $k_1$ and $k_2$,
respectively. Neglecting the electron mass,
the momenta, in the  $e^+e^-$ center-of-mass system, 
are given by:
\begin{eqnarray}
& & p_{1,2}=\frac{\sqrt{s}}{2} (1,0,0,\pm 1) \nonumber \\
& & k_{1,2}=\frac{\sqrt{s}}{2} (1\pm \frac{m_W^2 -m_{H^\pm}^2}{s},
\pm \kappa \sin\theta,0,\pm \kappa
\cos\theta), \nonumber
\end{eqnarray}
where $\sqrt{s}$ denotes the center of mass energy, 
$\theta$ the scattering angle
between $e^-$ and $W^+$, and $\kappa$ is determined by
$$\kappa ^2=(s-(m_{H^\pm} + m_W)^2)(s-(m_{H^\pm} -m_W)^2)/s^2 \, .$$
The Mandelstam variables are defined as follows:
\begin{eqnarray}
& & s =  (p_1+p_2)^2 = (k_1+k_2)^2  \nonumber\\
& & t = (p_1-k_1)^2 = (p_2-k_2)^2 =\frac{1}{2}(m_W^2  + m_{H^\pm}^2) - 
\frac{s}{2}
+\frac{s}{2} \kappa \cos\theta  \nonumber\\
& & u = (p_1-k_2)^2 = (p_2-k_1)^2 =\frac{1}{2} (m_W^2 + m_{H^\pm}^2) - 
\frac{s}{2}-
\frac{s}{2} \kappa \cos\theta \nonumber \\
& & s+t+u = m_W^2 + m_{H^\pm}^2 \, . \nonumber
\end{eqnarray}
 The differential cross section reads, expressed in terms of the
 one-loop amplitude ${\cal M}^1$: 
\begin{eqnarray}
\frac{d \sigma}{d \Omega}(e^+ e^- \to W^\pm H^\mp)= \frac{\kappa}{64 \pi^2 s} 
\sum_{Pol}
\vert {\cal M}^1 \vert^2 .
\end{eqnarray}
$ {\cal M}^1 $ can be decomposed into 
six invariant matrix elements ${\cal A}_i$ 
and scalar coefficient functions ${\cal M}_i$ as follows:
\begin{equation}
  {\cal M}^1 =\sum_{i=1}^{6} {\cal M}_i {\cal A}_i \, .
\label{decomp}
\end{equation}
The basic matrix elements ${\cal A}_i$ are given by the following expressions:
\begin{eqnarray}
& & {\cal A}_1 = \bar{v}(p_2) \not\epsilon (k_1) \frac{1+\gamma_5}{2} 
u(p_1)\nonumber \\ & &
{\cal A}_2 = \bar{v}(p_2) \not\epsilon (k_1) \frac{1-\gamma_5}{2} 
u(p_1)\nonumber \\ & &
{\cal A}_3 =\bar{v}(p_2) \not k_1 \frac{1+\gamma_5}{2} u(p_1) 
(p_1.\epsilon(k_1))\nonumber \\ & &
{\cal A}_4=\bar{v}(p_2) \not k_1 \frac{1-\gamma_5}{2} u(p_1) (p_1.\epsilon 
(k_1))\nonumber \\ & &
{\cal A}_5=\bar{v}(p_2) \not k_1 \frac{1+\gamma_5}{2} u(p_1) (p_2.\epsilon 
(k_1))\nonumber \\ & &
{\cal A}_6=\bar{v}(p_2) \not k_1 \frac{1-\gamma_5}{2} u(p_1) (p_2.\epsilon 
(k_1)) \, .
\label{inva}
\end{eqnarray}
The squared amplitude, after summation  
over the polarizations of the $W^+$ boson, 
gets the following form:
\begin{eqnarray}
& & \sum_{Pol} \vert {\cal M}^1 \vert^2 = 
2 s (\vert {{\cal M}_1 \vert }^2 + \vert{{\cal M}_2 \vert }^2) - 
\frac{(m_{H^\pm}^2m_W^2 - t u)}{4 m_W^2}
    \{ 4  \vert {{\cal M}_1 \vert }^2 + 4 \vert {{\cal M}_2 \vert }^2 
\nonumber \\ & &  + 4 (m_W^2 - t) Re[ 
{{\cal M}_1}{{\cal M}^*_3} +  {{\cal M}_2}{{\cal M}^*_4}]  +
        (m_W^2 - t )^2 [ \vert  {{\cal M}_3 \vert }^2 +  \vert {{\cal M}_4} 
\vert ^2]  \nonumber  \\ 
[0.2cm] & &
 +   4 (m_W^2 - u) Re[ {{\cal M}_1} {{\cal M}^*_5} + {{\cal M}_2} {{\cal M}^*_6}]
+  (m_W^2 - u)^2 [  \vert {{\cal M}_5 \vert }^2 +  \vert {{\cal M}_6 \vert }^2] 
\nonumber \\ [0.2cm] & &
 - 2 (m_{H^\pm}^2 m_W^2 + m_W^2 s - t u) Re[ {{\cal M}_3}  {{\cal M}^*_5} + 
 {{\cal M}_4}  {{\cal M}^*_6} ]  \}
\end{eqnarray}
The coefficient functions ${\cal M}_i$ can be read off from the 
explicit expressions in Appendix C, according to the decomposition
(\ref{decomp}).   

\renewcommand{\theequation}{4.\arabic{equation}}
\setcounter{equation}{0}
\section*{4. On-shell renormalization}
We have evaluated the  one-loop induced process
$e^+ e^- \to  W^\pm H^\mp$ in the 't Hooft - Feynman gauge, and using
dimensional regularization \cite{thooft}. 
The  types of Feynman diagrams are depicted
in Figure 1. It displays
the corrections to the $\gamma$--$W^\pm H^\mp$ and to 
the $Z$--$W^\pm H^\mp$ vertices, box diagrams, 
contributions coming from the various mixings  
 $H^-$--$W^+$  and  $H^-$--$G^+$ in t and s channels, 
and finally the counter-terms.
Note that the $s$-channel $H^-$--$W^+$ mixing 
(Fig. 1.16 ) vanishes for an on-shell transverse W gauge boson.  
There is no contribution from the  $W^+$--$G^-$ mixing
because the $\gamma$--$G^\pm H^\mp$ and Z--$G^\pm H^\mp$ vertices are  
absent at the tree level. 
All the Feynman 
diagrams have been generated and computed using FeynArts \cite{seep} and 
FeynCalc \cite{rolf} packages.
We also use  the fortran FF--package \cite{ff} in the numerical analysis.

Although the amplitude for our process vanishes at the tree level,
complications like tadpole contributions and 
vector boson--scalar mixings require a careful treatment of renormalization.
We adopt  the on-shell renormalization scheme of \cite{dabelstein},
for the Higgs sector, which is an extension of the on-shell scheme in
\cite{Hollik}. 
In this scheme,  field renormalization is performed in the 
manifest-symmetric version of the Lagrangian. A field 
renormalization constant $Z_{\Phi_{1,2}}$ is assigned to each 
Higgs doublet 
$\Phi_{1,2}$. The Higgs fields and vacuum expectation values
 $v_i$ are renormalized as follows:    
\begin{eqnarray}
& &\Phi_i \rightarrow (Z_{\Phi_i})^{1/2} \Phi_i \nonumber\\
& &v_i \rightarrow (Z_{\Phi_i})^{1/2} (v_i-\delta v_i) \, . 
\label{reno}
\end{eqnarray}
With these
substitutions in the Lagrangian~(\ref{covder}), expanding 
the renormalization constants $Z_i=1+\delta Z_i$ to 
the one-loop order,
we obtain all the counter-terms relevant for our process: 
a counter-term for the $W-H$ 2-point function, and counter-terms for
the $\gamma WH$ and $ZWH$ vertices, visualized   
in Fig.1.26 to Fig.1.28
($k_1$ denotes the momentum of the outgoing $W^+$),
\begin{eqnarray}
& &\delta [W_\nu^\pm H^\mp] \quad = - \, k_1^{\mu} \Delta \label{deltac} \\
& &\delta [A_\nu W_\mu^\pm H^\mp] = i e g_{\mu\nu} \Delta
\label{deltaa} \\
& &\delta [Z_\nu W_\mu^\pm H^\mp]\, = i e g_{\mu\nu}
\frac{s_W}{c_W}
\Delta  \label{deltab}
\end{eqnarray} 
with 
\begin{equation}
\label{deltadef}
   \Delta =\frac{\sin 2 \beta}{2} m_W
    [\frac{\delta v_1}{v_1} - \frac{\delta v_2}{v_2} 
     +\frac{1}{2} (\delta Z_{\Phi_2} -\delta Z_{\Phi_1} )] \, . 
\end{equation}
In the on-shell scheme, the
following renormalization conditions are imposed:
\begin{itemize}
\item  The renormalized tadpoles, i.e.~the sum of
      tadpole diagrams $T_{h,H}$ and tadpole counter-terms
      $\delta_{h,H}$ vanish:   
\[   T_{h} +\delta t_h=0, \quad T_H +\delta t_H=0 \, . \]
These conditions 
guarantee that $v_{1,2}$ appearing in the renormalized Lagrangian ${\cal L_R}$ 
are located at the minimum of the one-loop potential.
\item  The real part of the renormalized 
non-diagonal self-energy $\hat{\Sigma}_{H^- W^+}(k^2)$ 
vanishes for an on-shell charged Higgs boson: 
\begin{eqnarray}
& & Re \hat{\Sigma}_{H^- W^+} (m_{H^{\pm}}^2)  =0   
\label{WH0}
\end{eqnarray}
 (the  $\hat{}$ labels the renormalized quantity
which is given in Appendix C).
This renormalization condition 
determines the term $\Delta$, and consequently 
$\delta [A_\nu W_\mu^\pm H^\mp]$ 
and $\delta [Z_\nu W_\mu^\pm H^\mp]$ are also fixed.
The explicit expressions are given in Appendix C. 
\end{itemize} 
The last renormalization condition is sufficient 
to discard the real part of the $H^-$--$G^+$ mixing contribution
as well. 
Indeed,  using the 
 Slavnov--Taylor identity \cite{9607485}, \cite{9809324}   
$$ k^2 {\Sigma}_{H^-W^+}(k^2) -m_W {\Sigma}_{H^-G^+}(k^2)=0\ \quad
\mbox{at}\ \ \ k^2 = m_{H^\pm}^2 $$
valid also for the renormalized quantities,
together with eq.~(\ref{WH0}),  
it follows that
$$ Re\hat{\Sigma}_{H^-G^+}(m_{H^\pm}^2)=0 \, . $$
In particular, the Feynman diagrams depicted in Fig.~1.25 will not contribute
with the above renormalization conditions, being purely real valued.

The amplitudes corresponding to the 
two counter-terms (photon and Z boson exchanges) in
eqs.~(\ref{deltaa}, \ref{deltab}) contain only the 
matrix elements $ {\cal A}_1$ and ${\cal A}_2$  as follows:
\begin{eqnarray}
& & 
    {\cal M}_{CT}^{\gamma} = -\frac{e^2}{s}
\Delta   ( {\cal A}_1 + {\cal A}_2)\nonumber \\ & &
    {\cal M}_{CT}^{\gamma} = -\frac{e^2}{s-m_Z^2}
\frac{s_W}{c_W} \Delta   ( (g_A+g_V){\cal A}_2 +
 (g_V-g_A){\cal A}_1) 
\end{eqnarray}
with the electron--$Z$ coupling constants 
\begin{eqnarray}
g_V=\frac{1}{4 s_W c_W}(1-4 s_W^2)\, , \quad  g_A=\frac{1}{4 s_W c_W}
\label{gvga}
\end{eqnarray}
and $s_W \equiv \sin \theta_W, c_W \equiv \cos \theta_W$.
As a consequence,
only the form factors $ {\cal M}_1$ and ${\cal M}_2$ can contain UV 
divergences; each of the remaining form factors should be
UV finite by itself, which provides a good analytical check of
the calculations.

\renewcommand{\theequation}{5.\arabic{equation}}
\setcounter{equation}{0}
\section*{5. Numerical results and discussion}
The difference between the predicted cross sections
in Model-I and Model-II is essentially due to the fermion loops containing
the different Yukawa couplings.
For small values of $\tan\beta$
the cross sections are approximately the same for both models
of type I and II (Figure 2a).
For $\tan\beta > 10$ (Figure 2b),
 the cross section decreases to very small values
in Model-I,
corresponding to the Yukawa coupling proportional to $1/\tan\beta$.
In Model-II, the
cross section has a minimum for $\tan\beta\approx 30$ and 
increases again for large  $\tan\beta$, following the 
$b$ Yukawa coupling proportional to $\tan\beta$.

In the following more explicit discussion we consider the range
$ 0.5 \leq \tan\beta < 10 $, where both type I and II models give
practically the same numerical results. In this interval for $\tan\beta$
the total cross section 
varies from 7.5 fb to 0.012 fb,  for $\sqrt{s}=500 GeV$ and 
$m_{H^\pm}=220 GeV$.

We will take the following experimental input for the physical parameters 
\cite{databooklet}:
\begin{itemize}
\item the fine structure constant: $\alpha=\frac{e^2}{4\pi}=1/137.03598$.
\item the gauge boson masses: $M_Z=91.187\ GeV$ and  $M_W=80.41\ GeV$.
In the on--shell scheme we define $\sin^2 \theta_W$ by 
$\sin^2 \theta_W\equiv 1- \frac{M_W^2}{M_Z^2}$, so that it 
receives by definition no loop corrections.
\item the top and bottom quarks masses are taken to be: $m_t=175$ GeV and 
$m_b=4.5$ GeV.
\end{itemize}
Fig.3 shows the top-bottom contribution to the 
 integrated cross section as a 
function of the center of mass energy $\sqrt{s}$  for four values of 
$m_{H^\pm}$ and for two values of $\tan\beta$, $0.6$ and $1.6$.
It can be seen that for a 
small $\tan\beta$ (Fig.3.a) the top quark effect is enhanced. 
One can reach a cross section 
of 3.5 $fb$ for a charged Higgs mass of about 220 GeV.
The cross section is enhanced for $m_{H^\pm}=170$ GeV 
owing to the proximity of the normal threshold cut of the three-point
function at $m_{H^\pm}= m_t + m_b$. 
 
As $\tan\beta$ increases the top quark effect decreases, leading to almost
an order of magnitude suppression of the  cross section for $\tan\beta=1.6$ 
(Fig. 3.b). For large values of $\tan \beta (\simeq 50)$ the bottom quark
contribution becomes leading and of comparable magnitude to that of
the top quark in the small $\tan \beta$ region. 

For the general THDM we will present our numerical results
in the following three 
configurations (where all the masses are in GeV): \\
$\bullet$ $C_1$:  $m_{H^\pm}=220$, $m_{H}=180$, $m_h =90$, 
                  $m_A =80$, $\tan\alpha=1.4$, $\tan\beta=3.6$ \\
$\bullet$ $C_2$:  $m_{H^\pm}=300$, $m_{H}=280$, $m_h =120$, 
                  $m_A=220$, $\tan\alpha=2.4$, $\tan\beta=1.6$ \\
$\bullet$ $C_3$:  $m_{H^\pm}=400$, $m_{H}=380$, $m_h=120$, 
                  $m_A=370$, $\tan\alpha = 3$, $\tan\beta = 2$ \\

In those three cases the present experimental bound on $\delta\rho$ is 
respected. In Fig.4.a,b,c we show the total cross section 
(including  all virtual boson and fermion contributions) as a function of the
center of mass energy $\sqrt{s}$ for the three configurations 
listed above and for four different values of $\lambda_3$ which
satisfy the nominal unitarity constraints (\ref{UNITARITY})   
One can see from those curves that, for a fixed 
$\tan\beta$, the cross section increases with
$\lambda_3$. The effect of $\lambda_3$ arises
essentially via the trilinear vertices  $H^0H^+H^-$ and $h^0H^+H^-$.  
Note that values of $\lambda_3$ larger than the ones chosen in Fig.4.a,b,c  
start violating the unitarity constraints  
on the $H^0H^+H^-$ and $h^0H^+H^-$ couplings.

Fig.4.d shows the ($\lambda_3$ independent) box contributions to the 
cross section in the $C_{1,2,3}$ configurations.
This contribution remains generically suppressed 
(0.01--0.005 fb for $\sqrt{s}=$500--1000 GeV), even in the favourable
light Higgs mass configurations such as in case $C_1$.\\
  
Figures 5.a, 5.b and 5.c show the total
cross section as a function of 
$\tan\beta$ in the $C_{1,2,3}$ configurations with $\lambda_3= 0.1$, and 
for $\sqrt{s}=500$ Gev, 1 TeV and 1.5 TeV respectively.
One can see that the cross section reaches a minimum
for moderate values of $\tan\beta$ while it gets much larger (a few orders of 
magnitude) for both low and high values of this parameter. 
The first enhancement (small $\tan\beta$) is due to the top quark effect as 
discussed above, while the second enhancement comes from the effect of large 
$\tan\beta$ in $H^0H^+H^-$ and $h^0H^+H^-$ couplings.
With the parameter choice of Fig.5, the nominal unitarity condition 
(\ref{UNITARITY}) forbids $\tan \beta$ larger than $14$. One should, however, 
stress that a different choice of the Higgs masses and $\lambda_3$ closer to 
the MSSM configurations would suppress the Higgs sector effect in the large 
$\tan \beta$ regime, thus allowing even larger $\tan \beta$ values to be 
consistent with Eq.(\ref{UNITARITY}). In this
case large effects from the bottom quark sector become dominant.

In Fig.6 we show the dependence of the cross section
 on the charged Higgs mass for $\sqrt{s}=500$ Gev, 1 TeV 
and 1.5 TeV in the case where $\tan\beta=2$, $\tan\alpha=3$, 
$\lambda_3=0.1$,
$m_H=180$ GeV, $m_h=90$ and $m_A=80$ GeV. 
One notes the high sensitivity of the cross section  to the Higgs mass when
the latter is above the $W^- h$ and $b\bar{t}$ decay thresholds 
(which happen to roughly coincide owing to our input values, and 
correspond to the kink in the plots). Below this threshold, the cross-section
is much less sensitive to the Higgs mass, however, one should keep in mind
that this region corresponds to the opening of the charged Higgs pair 
production which yields a much more interesting event rate.  

We mention that our results are in qualitative agreement with
those of ref.\cite{chine}, inasmuch as we can compare. Note however that
we did not assume any Higgs mass rules, and also the
renormalization scheme in \cite{chine} was not explicitly defined.
Quite recently, after this work was completed, another paper
\cite{kanemura} appeared dealing with the same subject.

\renewcommand{\theequation}{5.\arabic{equation}}
\setcounter{equation}{0}
\section*{6. Conclusions}
To summarize, we have computed the associated production of a 
charged Higgs boson
with a gauge boson $W$ at high energy $e^+e^-$ collisions. 
The calculation is performed within dimensional regularization
 in the on shell scheme. The study is done in the
general two Higgs doublet model taking into account constraints on the
$\rho$ parameter and unitarity constraints on 
the trilinear Higgs couplings. We have shown that in the small $\tan \beta$
regime the top effect is enhanced leading to important cross sections 
(about 1 fb), while the leading contributions in the large $\tan \beta$ regime,
come from the trilinear $HH^\pm H^\mp $ and $hH^\pm H^\mp $ vertices.
If the charged Higgs is too heavy to be produced in pairs in $e^+ e^-$
future machines, the associated production with a $W$ boson would be the only  
means to look for direct evidence for it. The smallness of the cross section
would require, however, a very high luminosity option.

\newpage
\renewcommand{\theequation}{A.\arabic{equation}}
\setcounter{equation}{0}

\section*{Appendix A:}
In this section we list the Feynman rules for 
the 3-point vertices involving charged Higgs bosons,
in the general THDM.
\begin{eqnarray}
g_{H^0H^+H^-}& = &-i\frac{g}{m_W} [ \cos(\beta-\alpha)(m^2_{\Hpm}
- \frac{m^2_H}{2}) + \frac{\sin(\alpha+\beta)}{\sin 2\beta}\{
4\lambda_3 v^2 +\frac{1}{2}(m^2_H+m^2_h)\nonumber \\
&&-\frac{1}
{2 \sin 2\beta}(\sin 2\alpha + 2 \sin (\alpha-\beta)
\cos(\alpha +\beta))(m^2_H-m^2_h) \}] \label{A6} \\
g_{h^0H^+H^-}&=& -i\frac{g}{m_W} [\sin(\beta-\alpha)(m^2_{\Hpm}
- \frac{m^2_h}{2}) + \frac{\cos(\alpha+\beta)}{\sin 2\beta}\{
4\lambda_3 v^2 +\frac{1}{2}(m^2_H+m^2_h) \nonumber \\
&&-\frac{1}
{2 \sin 2\beta}(\sin 2\alpha + 2 \sin (\alpha+\beta)
\cos(\alpha -\beta))(m^2_H-m^2_h) \}]\label{A7} \\
g_{H^0 H^{\pm}G^{\mp}}&=&\frac{-i g \sin(\beta-\alpha) (m_{H^\pm}^2-m_H^2)}{2 m_W}\\
g_{h^0 H^{\pm}G^{\mp}}&=&\frac{i g \cos(\beta-\alpha) (m_{H^\pm}^2-m_h^2)}{2 m_W}\\
g_{A^0 H^{\pm}G^{\mp}}&=& 
\mp \frac{m_{H^\pm}^2-m_A^2}{v \sqrt{2}}\label{58} \\
g_{H^0 G^+G^-}&=& -im_H^2\frac{\cos (\beta-\alpha)}{\sqrt{2} v} \label{H1GPGM}  \\
g_{h^0 G^+G^-}&=& -im_h^2\frac{\sin (\beta-\alpha)}{\sqrt{2}v}   \label{H2GPGM}
\end{eqnarray}
The parameter $\lambda_3$ enters only $g_{H^0H^+H^-}$ and  
$g_{h^0H^+H^-}$, while the vertices 
$g_{H^0 H^{\pm}G^{\mp}}, g_{h^0 H^{\pm}G^{\mp}}$
and $g_{A^0 H^{\pm}G^{\mp}}$ have a particularly simple form,
proportional to $m_H^2$ and $m_h^2$, respectively.

\renewcommand{\theequation}{B.\arabic{equation}}
\setcounter{equation}{0}

\subsection*{Appendix B: One-loop functions}
Let us briefly recall the definitions of 
scalar and tensor integrals \cite{pv}  we use.
The inverse of the propagators are denoted by
\begin{eqnarray} 
d_0= q^2-m_0^2 \ , \ d_i= (q+p_i)^2-m_i^2 \nonumber
\end{eqnarray}
where the $p_i$ are the momenta of the external particles (always incoming).\\
\\
{\bf  One-point function:}
\begin{eqnarray} 
A_0(m_0^2)=\frac{(2\pi \mu)^{(4-D)} }{i \pi^2}\int d^Dq \frac{1}{d_0} \nonumber
\end{eqnarray}
where $\mu$ is an arbitrary renormalization scale 
and $D$ is the space-time dimension.\\

\noindent
{\bf  Two-point functions:} \\

\begin{eqnarray} 
 B_{0,\mu} = \frac{(2\pi \mu)^{(4-D)} }{i \pi^2}\int 
d^Dq \frac{\{1,q_{\mu}\} }{d_0 d_1} 
\nonumber 
\end{eqnarray}
Using Lorentz  covariance, one gets for the vector integral 
\begin{eqnarray} 
 B_\mu  =  {p_1}_{\mu} B_1 \nonumber 
\end{eqnarray}
with the scalar function 
$B_1(p_1^2,m_0^2,m_1^2)$.  \\
\\
{\bf  Three-point functions:}
\begin{eqnarray} 
 C_{0,\mu,\mu \nu}  = 
\frac{(2\pi \mu)^{(4-D)} }{i \pi^2}
\int d^Dq \frac{ \{1, q_{\mu},q_{\mu}q_{\nu}\} }{d_0 d_1 d_2} \nonumber
\end{eqnarray}
where $p_{ij}^2=(p_i + p_j)^2$. 
 Lorentz covariance yields the decomposition
\begin{eqnarray}
& & C_\mu  =  {p_{1\mu} } C_1 +  p_{2\mu} C_2 \\ [0.4 cm]
& & C_{\mu\nu}  =  g_{\mu\nu} C_{00} + p_{1\mu} p_{1\nu} C_{11}
+ p_{2\mu} p_{2_\nu} C_{22} +
(p_{1\mu} p_{2_\nu} +p_{2\mu} p_{1_\nu} ) C_{12}
\end{eqnarray}
with the  scalar functions 
$C_{i,ij}(p_1^2,p_{12}^2,p_2^2,m_0^2,m_1^2,m_2^2)$.  \\
\\
{\bf  Four-point functions:}
\begin{eqnarray}
 D_{0,\mu,\mu \nu} =
\frac{(2\pi \mu)^{(4-D)} }{i \pi^2}
\int d^D q \frac{ \{1, q_{\mu},q_{\mu}q_{\nu}\} }{d_0 d_1 d_2 d_3} 
\end{eqnarray}
Again,  Lorentz covariance allows the decomposition
\begin{eqnarray}
& & D_\mu  =  {p_1}_{\mu} D_1 +  p_{2\mu} D_2 +  p_{3\mu} D_3 \\ [0.4 cm]
& & D_{\mu\nu}  =  g_{\mu\nu} D_{00} + p_{1\mu} p_{1\nu} D_{11}
+ p_{2\mu} p_{2\nu} D_{22} + 
p_{3\mu} p_{3\nu} D_{33} +
 (p_{1\mu} p_{2\nu} +p_{2\mu} p_{1\nu}) D_{12}+ \nonumber \\ [0.4 cm] & &
( p_{1\mu} p_{3\nu} +p_{3\mu} p_{1\nu}) D_{13} +
(p_{3\mu} p_{2\nu} + p_{2\mu} p_{3\nu}) D_{23}.
\end{eqnarray}
with the scalar 4-point functions
$D_{i,ij} (p_1^2,p_{12}^2,p_{23}^2,p_3^2,p_2^2,p_{13}^2,m_0^2,
m_1^2,m_2^2,m_3^2)$.

All the tensor coefficients 
can be algebraically reduced to the basic scalar integrals 
$A_0$, $B_0$, $C_0$ and $D_0$.
The analytical expression of all the scalar functions can be found in 
\cite{pv,ff}. 

\renewcommand{\theequation}{C.\arabic{equation}}
\setcounter{equation}{0}

\subsection*{Appendix C:  One-loop amplitude}
In this appendix, 
 we use the fermion-vector boson coupling constants 
as defined in terms of the neutral-current
and charged-current vertices
\begin{eqnarray}
& & [f\bar{f} V_{\mu}] =\gamma_{\mu} (g_V^{f L} \frac{1-\gamma_5}{2} 
+g_V^{f R} \frac{1+\gamma_5}{2}) \nonumber \\
& & [u\bar{d} W^+_{\mu}] = -\frac{1}{\sqrt{2} s_W} \gamma_{\mu} 
(\frac{1-\gamma_5}{2})
\end{eqnarray}
in the following notation:
\begin{eqnarray}
& &   g_{\gamma}^{f L}= g_{\gamma}^{f R}=-e_f  \nonumber \\
& &  g_{Z}^{uL}= -\frac{(1- 2 s_W^2 e_u)}{4 s_W c_W}\, , \qquad  
g_{Z}^{f R}= \frac{2s_W^2 e_f}{4 s_W c_W} \nonumber \\
& &  g_{Z}^{dL}= \frac{(1+ 2 s_W^2 e_d)}{4 s_W c_W}  \nonumber \\
& &e_u= -2 e_d= \frac{2}{3} |e| \, ,  \qquad e_e= - |e| \, . 
\end{eqnarray}
\\
\\
{\Large{\bf Vertex diagrams}}\\
\\
{\large{\bf Fermionic loops}}\\
\\
{\bf d-d-u exchange}\\
\\
The diagram with the $ddu$ triangle, Fig.1.1, yields the following
contribution to the one-loop amplitude (\ref{decomp}) for 
each $V$ boson exchange ($V=\gamma,Z$): 
\begin{eqnarray}
M_{1.1} & = & \frac{2 N_C \alpha^2}{(\sqrt{2}    (s-m_V^2 ) s_W)} \Bigm(
 \{ - (m_u\  Y_{ud}^L\ g_V^{d L} + m_d\  Y_{ud}^R\ (g_V^{d L} - g_V^{d R})) 
        B_0(s, m_d^2, m_d^2) - \nonumber \\ [0.2cm] & &
m_u  (m_u^2\  Y_{ud}^L\ g_V^{d L} + m_d\ m_u\  Y_{ud}^R\ (g_V^{d L} - g_V^{d R})  
          - m_d^2  Y_{ud}^L\  g_V^{d R} )  C_0   + \nonumber \\ [0.2cm] & &
       2 (m_u\  Y_{ud}^L + m_d\  Y_{ud}^R) g_V^{d L} 
  C_{00} \} [ g_V^{eR}\ {\cal A}_1 + g_V^{eL}\ {\cal A}_2 ] \nonumber \\ [0.2cm] & &
 - \{ (m_u\ (m_W^2 + u)\  Y_{ud}^L\ g_V^{d L} + m_d\  Y_{ud}^R\ (m_W^2\ g_V^{d L} -
 t\ g_V^{d R})) C_{1}  + m_u\  u\  Y_{ud}^L\ g_V^{d L} C_0   \nonumber \\ [0.2cm] & &
     +  (m_u\ (m_{H\pm}^2 +  u)\  Y_{ud}^L\ g_V^{d L} + 
          m_d\  Y_{ud}^R\ (u\ g_V^{d L}  -m_{H\pm}^2\ g_V^{d R})) C_2\}\ 
 g_V^{eR} {\cal A}_1 + \nonumber \\ [0.2cm] & &
\{ [ m_u\ (m_W^2 + t)\  Y_{ud}^L\ g_V^{d L} + m_d \ Y_{ud}^R \ (m_W^2 \ 
g_V^{d L} - u \ g_V^{d R})] 
        C_{1}  - m_u\  t\  Y_{ud}^L\ g_V^{d L} C_0   + \nonumber \\ [0.2cm] & &
       [m_u\ (-m_{H\pm}^2  -  t ) \ Y_{ud}^L\ g_V^{d L} + 
          m_d\  Y_{ud}^R\ (m_{H\pm}^2\  g_V^{d R} - t\ g_V^{d L} )] 
        C_{2} \}  g_V^{eL}\ {\cal A}_2  \nonumber \\ [0.2cm] & &
 - 2 \{ g_V^{d L} \ (m_u \ Y_{ud}^L \ C_0 +  m_u \ Y_{ud}^L \ C_{1}  + 
       (2 m_u\  Y_{ud}^L + m_d \ Y_{ud}^R) C_{2} +  \nonumber \\ [0.2cm] & &
       (m_u\  Y_{ud}^L + m_d \ Y_{ud}^R) C_{12}  + 
       (m_u\  Y_{ud}^L + m_d \ Y_{ud}^R) C_{22} )\} [  g_V^{eR} {\cal A}_3 +
 g_V^{eL} {\cal A}_6 ]  \nonumber \\ [0.2cm] & &
 + 2 \{  m_d \ Y_{ud}^R \ g_V^{d R}\  C_{1}  - 
       g_V^{d L} (m_u\  Y_{ud}^L\  C_{2}  + \nonumber \\ [0.2cm] & &
     (m_u\  Y_{ud}^L + m_d\  Y_{ud}^R) ( C_{12}  + 
 C_{22} )) \} [ g_V^{eL} {\cal A}_4 + g_V^{eR}{\cal A}_5 ] \Bigm ) \, .
\end{eqnarray}
 Therein, all the $C_i$ and $C_{ij}$ have the same arguments: 
$(m_W^2, s, m_{H\pm}^2, m_u^2, m_d^2, d_d^2)$.
 Summation has to be performed over all fermion families; in 
practice only the third quark generation is relevant. 
\\
\\
{\bf u-u-d exchange}\\
\\
 The corresponding expression for the 
the diagram with the $uud$ triangle in Fig.1.2 
is obtained from the previous one by making the following replacements:
$$ Y_{ud}^L \longleftrightarrow Y_{ud}^R \ \ , \ \  m_u \longleftrightarrow m_d \ \ , 
\ \  g_V^{d L} \longleftrightarrow g_V^{u L}
\ \ , \ \ g_V^{d R} \longleftrightarrow g_V^{u R} \ \ \mbox{and} \ \ t 
\longleftrightarrow u $$
and also ${\cal A}_3 \longleftrightarrow {\cal A}_5 $ and $ {\cal A}_6 
\longleftrightarrow {\cal A}_4 $.
\\
\\
{\large {\bf Bosonic Loops}}\\
\\
 Listed are always the analytic expression for each generic
diagram of Fig.1. 
The sum over all the particle contents, given in the
associated tables, yields the corresponding 
contribution to the one-loop the matrix element (\ref{decomp}).   
\subsubsection*{Fig.1.3}
\begin{eqnarray}
M_{1.3} & = & - \frac{\alpha^2 g_{VW^+W^-}}{s-m_V^2} g_{W^+W^- S} g_{H^-W^+ S}
\{ \Bigm (-B_0(s, m_W^2, m_W^2] - (m_S^2 - m_W^2 + s)
        C_0 +\nonumber \\ [0.2cm] & & 
       (m_{H\pm}^2 - m_W^2 - s) C_1  + 
       (m_{H\pm}^2 - m_W^2 + s) C_2 + 
       C_{0 0} \Bigm ) [ g_V^{eR}\ {\cal A}_1 +  g_V^{eL}\ {\cal A}_2 ]
\nonumber \\ [0.2cm] & & 
  +   (4 C_1 +  C_2 - C_{1 2} -  C_{2 2} ) 
   [ g_V^{eR}\ ( {\cal A}_3 + {\cal A}_5) + 
    g_V^{eL}\ ({\cal A}_6 + {\cal A}_4) ] \} .
\end{eqnarray}
All the $C_i$ and $C_{ij}$ have the same arguments:
$(m_W^2, s, m_{H\pm}^2, m_{S}^2, m_{W}^2, m_{W}^2)$.
The couplings are summarized in the following table;
 $S$ is a generic notation for one of the Higgs bosons. 
\begin{center}
\begin{tabular}{|c|c|c|c|c|c|}  \hline\hline
 $S$  & $g_{ZZS}$   & $g_{W^+W^-S}$ & $g_{H^-W^+ S} $ & $g_{ZW^+W^-} $ & 
$g_{\gamma W^+W^-} $   \\ \hline \hline
 $h$  &$m_Z\frac{s_{\beta\alpha}}{c_W s_W }$  &$m_W\frac{s_{\beta\alpha}}{ s_W }$  
& $\frac{c_{\beta\alpha}}{ 2
s_W}$   & $-\frac{c_w}{s_W}$ & -1 \\ \hline
$H$ & $m_Z\frac{c_{\beta\alpha}}{c_W s_W}$ & $m_W\frac{c_{\beta\alpha}}{ s_W}$ & 
$\frac{-s_{\beta\alpha}}{ 2 s_W} $  &  $-\frac{c_w}{s_W}$ & -1\\ \hline
\end{tabular} \end{center}
Therein, $ s_{\beta \alpha} \equiv \sin (\beta - \alpha)$ 
and $c_{\beta \alpha} \equiv \cos (\beta - \alpha) $. 

\subsubsection*{Fig.1.4}
\begin{eqnarray}
M_{1.4}& = & \frac{2\alpha^2 }{s-m_V^2}g_{VS_iS_j} g_{W^-W^+S_j} g_{H^-W^+S_i}\{
     C_{0 0} [ g_V^{eR}\ {\cal A}_1 +  g_V^{eL}\ {\cal A}_2 ] \nonumber \\ [0.2cm]
& & - (2 C_0 + 2 C_{1} + 3 C_{2} + C_{1 2} +  C_{22})
 [ g_V^{eR}\ ( {\cal A}_3 + {\cal A}_5) + 
g_V^{eL}\ ({\cal A}_6 + {\cal A}_4) ] \}
\end{eqnarray}
where the arguments of $C_i$ and $C_{ij}$ are now as follows:
$(m_W^2, s, m_{H\pm}^2, m_{W}^2, m_{S_i}^2, m_{S_j}^2)$.
The couplings are summarized in the following table:
\begin{center}
\begin{tabular}{|c|c|c|c|}  \hline\hline
 ($S_i$,$S_j$ )    &  $g_{Z S_i S_j} $ 
& $g_{H^{-}S_i W }$
&$g_{W^-W^+S_j}$\\ \hline \hline
 ($A_0$, $h$)    & $\frac{c_{\beta\alpha}}{2 s_W c_W}$   &
$ \frac{1}{2 s_W} $ & $ s_{\beta\alpha}\frac{m_W}{s_W} $  \\ \hline
 ($A_0$, $H$)   & $-\frac{s_{\beta\alpha}}{2 s_W c_W}  $   &
$  \frac{1}{2 s_W} $ & $ c_{\beta\alpha}\frac{m_W}{s_W} $    \\ \hline
\end{tabular} \end{center}

\subsubsection*{Fig.1.5}
\begin{eqnarray}
M_{1.5}& = & \frac{\alpha^2 }{s-m_V^2}g_{VS_iS_j} g_{W^+S_jS_k} g_{H^-S_iS_k}\{
     C_{0 0} [ g_V^{eR}\ {\cal A}_1 +  g_V^{eL}\ {\cal A}_2 ] - \nonumber \\ [0.2cm]
& &
(C_{2} + C_{1 2} +  C_{2 2}) [ g_V^{eR}\ ( {\cal A}_3 + {\cal A}_5) + 
g_V^{eL}\ ({\cal A}_6 + {\cal A}_4) ] \}
\end{eqnarray}
All the $C_i$ and $C_{ij}$ have the same arguments:
$(m_W^2, s, m_{H\pm}^2, m_{S_k}^2, m_{S_i}^2, m_{S_j}^2)$.
The couplings are summarized in the following table:
\begin{center}
\begin{tabular}{|c|c|c|c|c|}  \hline\hline
 ($S_i$,$S_j$,$S_k$ )    & $g_{\gamma S_i S_j} $  & $g_{Z S_i S_j} $ 
& $g_{H^{\pm}S_iS_k}$
&$g_{W^+S_jS_k}$\\ \hline \hline
 ($A_0$, $h$, $G^\pm$)    & 0  & $i\frac{c_{\beta\alpha}}{2 s_W c_W}$   &
$ ig_{H^- A_0 G^+}$ & $ \frac{s_{\beta\alpha}}{2 s_W} $  \\ \hline
 ($A_0$, $H$, $G^\pm$ )   & 0 &  $-i\frac{s_{\beta\alpha}}{2 s_W c_W}  $   &
$  ig_{H^+ A_0 G^{-} } $ & $ \frac{c_{\beta\alpha}}{2 s_W} $    \\ \hline
 ($G^\pm$, $G^\pm$, $h$ )   & -1 &  $-\frac{\cos(2\theta_W)}{2 s_W c_W} $   &
$ g_{H^- h G^+} $  & $ \frac{s_{\beta\alpha}}{2 s_W} $  \\ \hline
 ($G^\pm$, $G^\pm$, $H$ )   & -1 & $-\frac{\cos(2\theta_W)}{2 s_W c_W} $   &
$ g_{H^- H G^+} $  & $ \frac{c_{\beta\alpha}}{2 s_W} $  \\ \hline
 ($H^\pm$, $H^\pm$, $h$ )   & -1 & $ -\frac{\cos(2\theta_W)}{2 s_W c_W} $   &
$ g_{H^- h H^+} $  &  $ \frac{c_{\beta\alpha}}{2 s_W} $  \\ \hline
 ($H^\pm$, $H^\pm$, $H$ )   & -1 & $ -\frac{\cos(2\theta_W)}{2 s_W c_W} $   &
$ g_{H^- H H^+} $  &  $ -\frac{s_{\beta\alpha}}{2 s_W} $  \\ \hline
\end{tabular} \end{center}
where the $g_{H^{\pm}S_iS_k}$ couplings have been defined in appendix A.

\subsubsection*{Fig.1.6}
\begin{eqnarray}
M_{1.6} & = & -\frac{\alpha^2 }{s-m_V^2}g_{V V' S_i} g_{W^+V'S_j} g_{H^-S_iS_j}
 C_0 [ g_V^{eR}\ {\cal A}_1 +  g_V^{eL}\ {\cal A}_2 ]
\end{eqnarray}
 where  
$ C_0= C_0(m_W^2, s, m_{H\pm}^2, m_{S_i}^2, m_{S_j}^2, m_{V'}^2)$.
The couplings are summarized in the following table:

\begin{center}
\begin{tabular}{|c|c|c|c|c|}  \hline\hline
 ($S_i$,$S_j$,$V'$ )    & $g_{\gamma V' S_i } $  & $g_{Z V' S_i} $ 
& $g_{H^{+}S_iS_j}$
&$g_{W^+V'S_j}$\\ \hline \hline
 ($h$, $G^\pm$, Z )    & 0 & $m_Z\frac{s_{\beta\alpha}}{s_W c_W}$   &
$ g_{H^-G^+ h}$ & $ -m_Z s_W  $  \\ \hline
 ($H$, $G^\pm$, Z )   & 0 &  $m_Z\frac{c_{\beta\alpha}}{s_W c_W}  $   &
$  g_{ H^{-} G^+ H} $ & $ -m_Z s_W $    \\ \hline
 ($G^\pm$, $h$, $W^\pm$ )   & $m_W $ &  $-m_Z s_W $   &
$ g_{H^- G^+h} $  & $ s_{\beta\alpha}\frac{m_W}{ s_W} $  \\ \hline
 ($G^\pm$, $H$, $W^\pm$ )   & $m_W$ &  $- m_Z s_W $   &
$ g_{H^-G^+H} $  & $ c_{\beta\alpha}\frac{m_W}{ s_W} $  \\ \hline
\end{tabular} \end{center}

\subsubsection*{Fig.1.7}

\begin{eqnarray}
M_{1.7}& = & \frac{2\alpha^2 }{s-m_V^2}g_{V V' S_i} g_{W^+S_iS_j} g_{H^-V'S_j}\{
-C_{00} [ g_V^{eR}\ {\cal A}_1 +  g_V^{eL}\ {\cal A}_2 ]
 +   \nonumber \\ [0.2cm] & &   (- C_{2} + 
        C_{1 2} +
        C_{22}) [ g_V^{eR}\ ( {\cal A}_3 + {\cal A}_5) + 
g_V^{eL}\ ({\cal A}_6 + {\cal A}_4) ]\}
\end{eqnarray}
All the $C_i$ and $C_{ij}$ have the same arguments:
$(m_W^2, s, m_{H\pm}^2, m_{S_i}^2, m_{S_j}^2, m_{V'}^2)$.
The couplings are summarized in the following table:
\begin{center}
\begin{tabular}{|c|c|c|c|c|}  \hline
\hline
 ($S_i$,$S_j$ )    &  $g_{\gamma W^+ S_i} $ &  $g_{Z W^+ S_i} $
& $g_{H^{-}W^+ S_j}$ &$g_{W^+S_iS_j}$\\ \hline \hline
 ($G^\pm$, $h$)  & $ m_w $   & $ -m_Z s_W$   &
$ \frac{c_{\beta\alpha} }{2 s_W} $ & $ \frac{s_{\beta\alpha}}{2 s_W} $  \\ \hline
 ($G^\pm$, $H$)  &  $ m_W$  & $-m_Z s_W $   &
$  -\frac{s_{\beta\alpha}}{2 s_W} $ & $ \frac{c_{\beta\alpha}}{2s_W} $    \\ \hline
\end{tabular} \end{center}

\subsubsection*{Fig.1.8}

\begin{eqnarray}
M_{1.8}& = & -\frac{\alpha^2 }{s-m_Z^2}g_{Z Z S} g_{ZW^-W^+} g_{H^-W^+S}\{
 ( B_0(s, m_S^2, m_Z^2) +  (2 m_{H\pm}^2 + 3 m_W^2 - 2s)
        C_0\nonumber \\ & & + [m_{H\pm}^2 + 3 m_W^2 - s] C_{1} +
 [3 m_{H\pm}^2 + m_W^2 - s] C_{2} - 
       C_{0 0}) [ g_V^{eR}\ {\cal A}_1 +  g_V^{eL}\ {\cal A}_2 ]\nonumber \\ & &
(2 C_0 - 2 C_{1} + 3 C_{2} + C_{1 2}+ C_{22})[ g_V^{eR}\ ( {\cal A}_3 + {\cal A}_5) + 
g_V^{eL}\ ({\cal A}_6 + {\cal A}_4) ] \}
\end{eqnarray}
where all the $C_i$ and $C_{ij}$ have the same arguments:
$(m_W^2, s, m_{H\pm}^2, m_{W}^2, m_{Z}^2, m_{S}^2)$.
The couplings are summarized in the following table:
\begin{center}
\begin{tabular}{|c|c|c|c|}  \hline\hline
 $S$    & $g_{ZZS}$ & $g_{H^-W^+ S} $ & $g_{ZW^+W^-} $   \\ \hline \hline
 $h$    &$m_Z\frac{s_{\beta\alpha}}{ s_W c_W}$  & $\frac{c_{\beta\alpha}}{ 2
s_W}$   & $-\frac{c_w}{s_W}$ \\ \hline
$H$ & $m_Z\frac{c_{\beta\alpha}}{ s_W c_W}$ & 
$\frac{-s_{\beta\alpha}}{ 2 s_W} $  &  $-\frac{c_w}{s_W}$\\ \hline
\end{tabular} \end{center}

\subsubsection*{Fig.1.9}

\begin{eqnarray}
M_{1.9} & = & \frac{\alpha^2 }{s-m_V^2} g_{VW^+H^-S_1} g_{W^+W^-S_1} 
\{B_0(m_W^2, m_{S_1}^2, m_W^2) [ g_V^{eR}\ {\cal A}_1 +  g_V^{eL}\ {\cal A}_2 ]\}
\end{eqnarray}

\subsubsection*{Fig.1.10}

\begin{eqnarray}
M_{1.10} & = & -\frac{\alpha^2 }{s-m_V^2} g_{VW^+S_1S_2} g_{H^-S_1S_2} 
\{ B_0(m_{H\pm}^2, m_S^2, m_{H\pm}^2) [ g_V^{eR}\ {\cal A}_1 +  g_V^{eL}\ {\cal A}_2 ]\}
\end{eqnarray}

\subsubsection*{Fig.1.11}

\begin{eqnarray}
M_{1.11} & = & \frac{\alpha^2 }{s-m_Z^2} g_{ZW^+H^-S_1} g_{SZZ} 
\{B_0(s, m_{S_1}^2, m_Z^2) [ g_V^{eR}\ {\cal A}_1 + g_V^{eL}\ {\cal A}_2 ]\}
\end{eqnarray}

where the couplings are given in the table

\begin{center}
\begin{tabular}{|c|c|c|c|}  \hline\hline
 $(S_1,S_2)$  & $g_{ZW^+S_1 S_2}$   &$g_{\gamma W^+S_1 S_2}$  & $g_{H^-S_1 S_2} $ 
   \\ \hline \hline
 $(h,H^-)$  & $-\frac{c_{\beta\alpha}}{2 c_W} $ 
  & $ \frac{c_{\beta\alpha}}{2 s_W} $  & $ g_{hH^-H^-} $ 
  \\ \hline
 $(H,H^-)$  & $\frac{s_{\beta\alpha}}{2 c_W} $ 
  & $ -\frac{s_{\beta\alpha}}{2 s_W} $  &  $g_{HH^-H^-} $ 
  \\ \hline
 $(h,G^+)$  & $-\frac{s_{\beta\alpha}}{2 c_W} $ 
  & $ \frac{s_{\beta\alpha}}{2 s_W} $  &  $g_{hH^-G^+} $ 
  \\ \hline
 $(H,G^+)$  & $-\frac{c_{\beta\alpha}}{2 c_W} $ 
  & $ \frac{s_{\beta\alpha}}{2 s_W} $  &  $g_{HH^-G^+} $ 
  \\ \hline
\end{tabular} \end{center}
\noindent
\\
{\Large{\bf Box diagrams}}
\subsubsection*{Fig.1.12}
\begin{eqnarray}
M_{h}^{Box} &= &- \frac{\alpha^2}{4 s_W^2} c_{\beta\alpha}s_{\beta\alpha}m_W
      \{ [C_0(m_e^2,m_e^2,s,m_W^2,0,m_W^2) + (m_h^2 - u) D_0 + (m_W^2 - u)
D_1  \nonumber\\ & &
    + (m_W^2 - s - t) D_3 ] {\cal A}_2 - 4 D_1 {\cal A}_4 \}
\end{eqnarray}
where all the $D_i$  functions   have as arguments: 
$(m_W^2,m_e^2, m_e^2, m_{H\pm}^2,s,m_h^2, m_W^2,0,m_W^2)$.
 The corresponding box graph with a heavy neutral scalar is obtained 
by the substitution 
\begin{eqnarray}
M_{H}^{Box} &= & - M_{h}^{Box} (m_h\to m_H)
\end{eqnarray} 

\subsection*{Counter-term diagrams}
 Denoting by 
  $k^{\mu} \,\Sigma_{H W}(k^2)$ 
the bare $H^\pm W^\pm$ 2-point  function,
Fig.1.17 - Fig.1.20,
the renormalized one is obtained by adding the counter-term
(\ref{deltac}). The renormalized  
non-diagonal self-energy 
$\hat{\Sigma}_{HW}(k^2)$ is thus given by:
\begin{eqnarray}
\hat{\Sigma}_{HW}(k^2) = \Sigma_{HW}(k^2) - \Delta \, .
\end{eqnarray}
 According to the condition (\ref{WH0}), 
requiring that the renormalized 
$W^+$--$H^-$ mixing vanishes, $\Delta$ is determined by: 
\begin{equation}
  \Delta =\mbox{Re} \, \Sigma_{HW}(m_{H\pm}^2 ) \, . 
\end{equation}
 The self-energy 
$\Sigma_{HW}(m_{H\pm}^2 )\equiv \Sigma_{HW}$ has the following
explicit form:
$$ \Sigma_{HW}= \Sigma_{HW}^{ferm}  +  \Sigma_{HW}^{bos} \, . $$
The fermionic part is the sum over the individual fermion families,
each one contributing 
\begin{eqnarray}
\Sigma_{HW}^{ud} & = & - N_C \frac{\alpha}{2\sqrt{2} s_W\pi}
     (m_d Y_{ud}^R B_0(m_{H^+}^2,m_d^2,m_u^2) 
      + (m_u Y_{ud}^L+m_d Y_{ud}^R) B_1(m_{H^+}^2,m_d^2,m_u^2) ) \, , 
    \nonumber \\
    & &  
\end{eqnarray}
and the bosonic part is given by 
\begin{eqnarray}
   \Sigma_{HW}^{bos}  & =& \frac{\alpha}{8 \pi s_W^2}
     ( c_{\beta\alpha} g_{hH^-H^-}  s_W B_0(m_{H\pm}^2,m_{h}^2,m_{H\pm}^2) - 
      c_{\beta\alpha} m_W s_{\beta\alpha} B_0(m_{H\pm}^2,m_{h}^2,m_W^2) +
     \nonumber \\
& &
     g_{hH^-G^+} s_{\beta\alpha} s_W B_0(m_{H\pm}^2,m_{h}^2,m_W^2) - 
      g_{HH^+H^-} s_{\beta\alpha} s_W B_0(m_{H\pm}^2,m_{H}^2,m_{H\pm}^2) +\nonumber \\
& & 
     c_{\beta\alpha} m_W s_{\beta\alpha} B_0(m_{H\pm}^2,m_{H}^2,m_W^2) + 
     c_{\beta\alpha} g_{HH^-G^+} s_W B_0(m_{H\pm}^2,m_{H}^2,m_W^2) + \nonumber \\
& &
     2 c_{\beta\alpha} g_{hH^+H^-} s_W B_1(m_{H\pm}^2,m_{h}^2,m_{H\pm}^2) + 
     c_{\beta\alpha} m_W s_{\beta\alpha} B_1(m_{H\pm}^2,m_{h}^2,m_W^2) + \nonumber \\
& &
     2 g_{hH^+G^+} s_{\beta\alpha} s_W B_1(m_{H\pm}^2,m_{h}^2,m_W^2) - 
     2 g_{H H^+H^-} s_{\beta\alpha} s_W B_1(m_{H\pm}^2,m_{H}^2,m_{H\pm}^2) - 
\nonumber \\
& &
      c_{\beta\alpha} m_W s_{\beta\alpha} B_1(m_{H\pm}^2,m_{H}^2,m_W^2) + 
     2 c_{\beta\alpha} g_{HH^+G^+} s_W B_1(m_{H\pm}^2,m_{H}^2,m_W^2) ) \, . 
\end{eqnarray}
Then the  amplitudes containing the counter terms, Fig.1.28, read: 
 
\begin{eqnarray}
M_{\delta (V W H)} =\frac{4 \pi \alpha g_V}{s-m_V^2} Re(\Sigma_{HW}) 
[ g_V^{eR}\ {\cal A}_1 +  g_V^{eL}\ {\cal A}_2 ]
\end{eqnarray}
with $ g_Z=-s_W/c_W, g_\gamma =-1 $ and  $g_V^{e L,R}$ 
defined in (C.1)

\subsection*{Amplitudes with non-diagonal self-energies}
\subsubsection*{t-channel $W^+$--$H^-$ mixing}
The generic diagram for this topology is depicted in Fig.1.13. 
The amplitude  contains only the invariant ${\cal A}_2$ :
\begin{eqnarray}
M_{1.13}=\frac{e^2}{2 s_W^2}\frac{ \mbox{Im}(\Sigma_{HW}^{ferm})
  + \mbox{Im}(\Sigma_{HW}^{bos}) }{m_{H^\pm}^2 -m_W^2} {\cal A}_2
\end{eqnarray}

\subsubsection*{s-channel $W^+$--$H^-$ mixing}
The generic diagram for this topology is depicted in Fig.1.15. 
The amplitude can be projected  on ${\cal A}_{1,2}$ as follows:
\begin{eqnarray}
M_{1.15}=\frac{e^2 g_{VWW} }{(s-m_V^2)}
(m_W^2- s)\frac{ \mbox{Im}(\Sigma_{HW}^{ferm})
  + \mbox{Im}(\Sigma_{HW}^{bos}) }{m_{H^\pm}^2 -m_W^2} (g_V^{eR}  {\cal A}_1 +
g_V^{eL} {\cal A}_2 )
\end{eqnarray}
where $g_{ZWW}=c_W/s_W$ and $g_{\gamma WW}=1$

\subsubsection*{s-channel $G^+$--$H^-$ mixing}
 The non-diagonal Goldstone--$H^\pm$ self-energy is
$\Sigma_{GH}$
evaluated in the same way 
as done before for the $W$-$H^\pm$ case
($k^2= m_{H^\pm}^2$):
$$ \Sigma_{GH} = \Sigma_{GH}^{ferm} + \Sigma_{GH}^{bos} \, , $$
where the fermionic part is obtained by summing over the families
with 
\begin{eqnarray}
\Sigma_{GH}^{ud}&=&
\frac{-\alpha N_C}{(2 \sqrt{2} m_W \pi s_W) }
\{ (m_u Y_{ud}^L - m_d Y_{ud}^R)  A_0(m_d^2) 
+ m_u(- m_d^2 +m_u^2) Y_{ud}^L  B_0(m_{H^\pm}^2 , m_d^2, m_u^2) 
        \nonumber \\ & &
       + m_{H^\pm}^2 ( m_u Y_{ud}^L -m_d Y_{ud}^R)   
        B_1(m_{H^\pm}^2 , m_u^2, m_d^2)\} \, ,
\end{eqnarray}
and the bosonic part is given by 
\begin{eqnarray}
& & \Sigma_{GH}^{bos}=
\frac{\alpha}{4\pi} \{
 g_{hH^+H^-}g_{h G^+H^-} B_0(m_{H^\pm}^2, m_h^2, m_{H^\pm}^2)
 +g_{HH^+H^-}g_{H G^+H^-}  B_0(m_{H^\pm}^2, m_H^2, m_{H^\pm}^2)\nonumber \\ & & +
  g_{hG^+H^-}g_{h G^+G^-}  B_0(m_{H^\pm}^2, m_h^2, m_W^2)
+ g_{HG^+H^-}g_{H G^+G^-} B_0(m_{H^\pm}^2, m_H^2, m_W^2)-\nonumber \\ & &
 \frac{s_{\beta\alpha} c_{\beta\alpha} }{4 s_W^2}
      (A_0(m_W^2) + (m_h^2 +m_{H^\pm}^2) B_0(m_{H^\pm}^2, m_h^2, m_W^2) - 
        2m_{H^\pm}^2  B_1(m_{H^\pm}^2, m_h^2, m_W^2) +\nonumber \\ & &
      - ( A_0(m_W^2) + (m_H^2 + m_{H^\pm}^2 ) B_0(m_{H^\pm}^2, m_H^2, m_W^2) - 
        2 m_{H^\pm}^2 B_1(m_{H^\pm}^2, m_H^2, m_W^2))  \} \, .
\end{eqnarray}
Then the amplitude of the diagram Fig.1.14 
 can be written in the form
\begin{eqnarray}
M_{1.14}=\frac{e^2 g_{VW^+G^-} }{(s-m_V^2)}
\frac{ \mbox{Im}(\Sigma_{GH}^{ud})
  + \mbox{Im}(\Sigma_{GH}^{bos}) }{m_{H^\pm}^2 -m_W^2} (g_V^{eR}  {\cal A}_1 +
g_V^{eL} {\cal A}_2 )
\end{eqnarray}
where $g_{ZW^+G^-}=-m_W s_W/c_W$ and $g_{\gamma W^+G^-}=m_W$.

\newpage

{\Large \bf Figure Captions:}\\
\begin{itemize}
\item[{\bf Fig. 1:}] Feynman diagrams relevant for the  
$e^+ e^- \rightarrow W^+ H^- $. Fig.1.1 $\to$ Fig.1.11 vertex diagrams, 
Fig.1.12 box diagram, Fig.1.13 $\to$ Fig.1.16
 typical diagrams contributing the self energies mixing
of $W^+$--$H^-$ and  $G^+$--$H^-$, 
self-energies of the mixing 
$W^+$--$H^-$ Fig.1.17 $\to$ Fig.1.20, 
self-energies of the mixing 
$G^+$--$H^-$ Fig.1.21 $\to$ Fig.1.25. 
 Fig.1.26 and  Fig.1.27 are the counter term for the mixing $G^+$--$H^-$ and 
 $W^+$--$H^-$. Fig. 1.28 are the counter--terms for the  
photon--$W^+$--$H^-$ and Z--$W^+$--$H^-$ vertices.

\item[{\bf Fig. 2:}] Top--bottom contribution to the 
integrated  cross section  as a function of 
$\tan\beta$ in Model type I and II
 for  $m_{H^\pm}=220 GeV$,  and $\sqrt{s}=500 GeV$.

\item[{\bf Fig. 3:}] Top--bottom contribution to the 
integrated  cross section  as a function of 
$\sqrt{s}$  
for four values of 
$m_{H^\pm}=140 GeV, \ 185 GeV, \ 300 GeV$,  and $\ 400 GeV$.
Fig. 3.a $\tan\beta=0.6$ and Fig. 3.b $\tan\beta=1.6$

\item[{\bf Fig. 4:}] Integrated total cross section  as a function of 
$\sqrt{s}$ for four values of $\lambda_3$.\\

Fig. 4.a $C_1$ case: $m_{H^\pm}=220$, $m_{H}=180$, $m_h =90$, $m_A =80$, $\tan\alpha=1.4$, $\tan\beta=3.6$  \\
Fig. 4.b $C_2$ case:  $m_{H^\pm}=300$, $m_{H}=280$, $m_h =120$, $m_A=220$, 
$\tan\alpha=2.4$, $\tan\beta=1.6$ \\
Fig. 4.c $C_3$ case:  $m_{H^\pm}=400$, $m_{H}=380$, $m_h=120$, $m_A=370$, $\tan\alpha = 3$, $\tan\beta = 2$ \\
(where all masses are in GeV)\\
Fig. 4.d Box contributions to the total cross section as function of the 
center of mass energy for $C_{1,2,3}$ configurations.

\item[{\bf Fig. 5:}] Integrated total cross section  as a function of 
$\tan\beta$  in the case of  the $C_{1,2,3}$ configurations  
with $\lambda_3=0.1$ and 
for $\sqrt{s}=500$ GeV (Fig5.a), 1 TeV (Fig.5.b) and 1.5 TeV (Fig.5.c) 
respectively.

\item[{\bf Fig. 6:}] Integrated total cross section  as a function of 
$m_{H^\pm}$ for $\sqrt{s}= 500,\ 1000,$ and $1500$ GeV,
with $\tan\beta=2$, $\tan\alpha=3$, $\lambda_3=0.1$,
$m_H=180$ GeV, $m_h=90$GeV and $m_A=80$ GeV

\end{itemize}

\newpage
\renewcommand{\thepage}{}
\begin{minipage}[t]{19.cm}
\setlength{\unitlength}{1.in}
\begin{picture}(0.1,0.1)(0.8,8.2)
\centerline{\epsffile{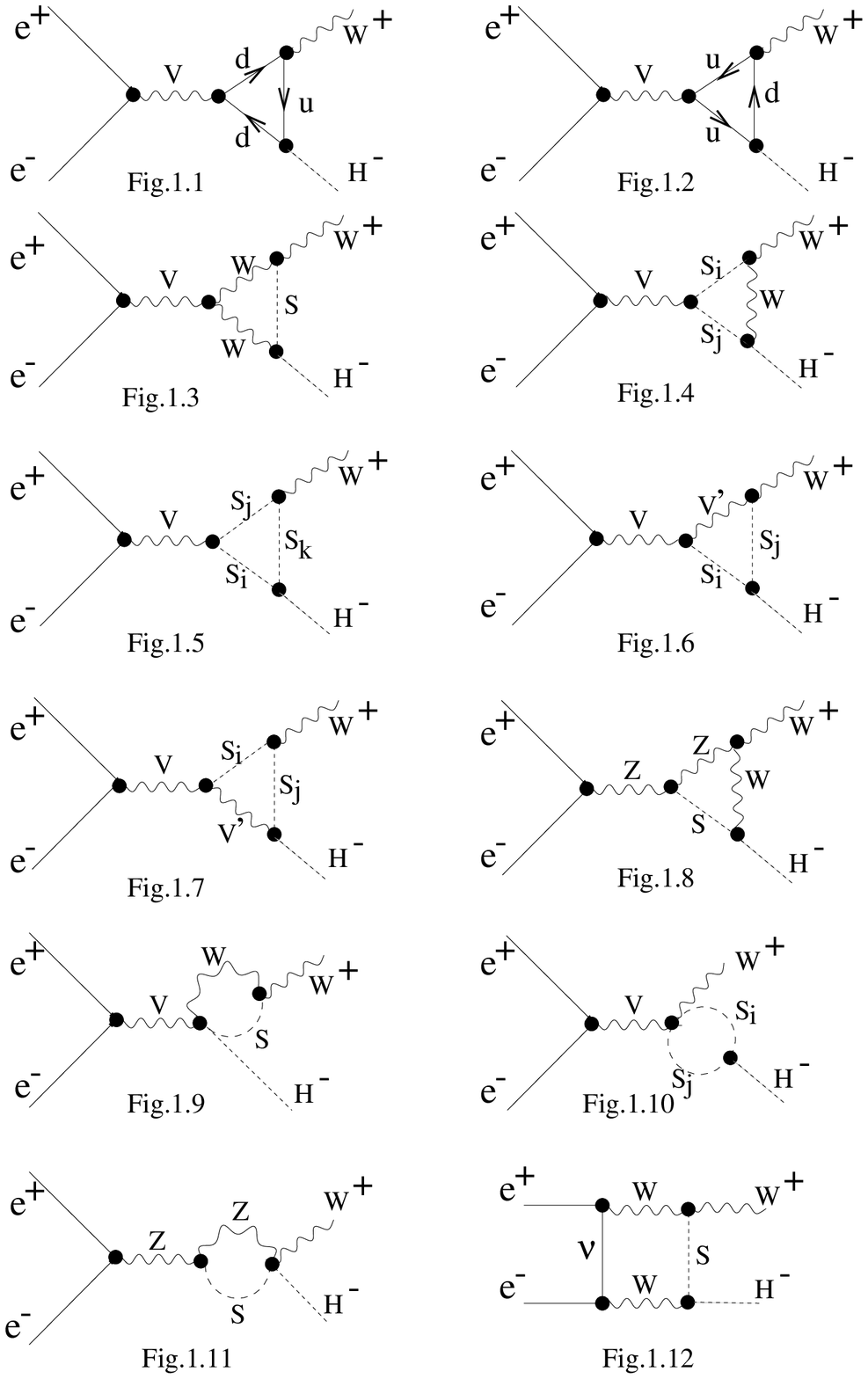}}
\end{picture}
\vspace{21.cm}

\hspace{7.cm}{\bf Figure. 1}
\end{minipage}

\newpage
\begin{minipage}[t]{19.cm}
\setlength{\unitlength}{1.in}
\begin{picture}(0.1,0.1)(0.8,7.9)
\centerline{\epsffile{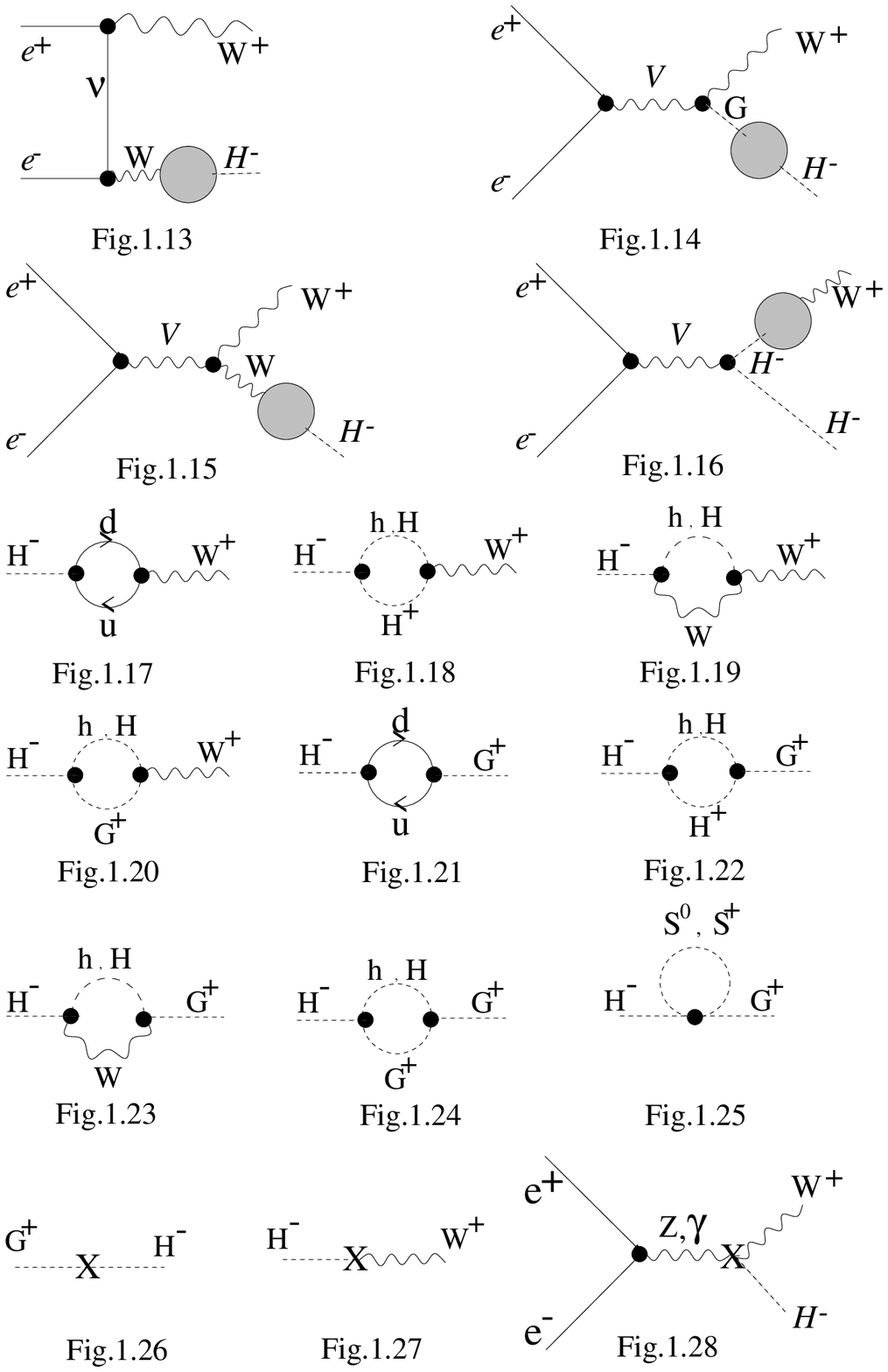}}
\end{picture}
\vspace{21.cm}

\hspace{7.cm}{\bf Figure.1 (cont.)}
\end{minipage}

\newpage
\begin{minipage}[t]{19.cm}
\setlength{\unitlength}{1.in}
\begin{picture}(1,1)(1.,9.1)
\centerline{\epsffile{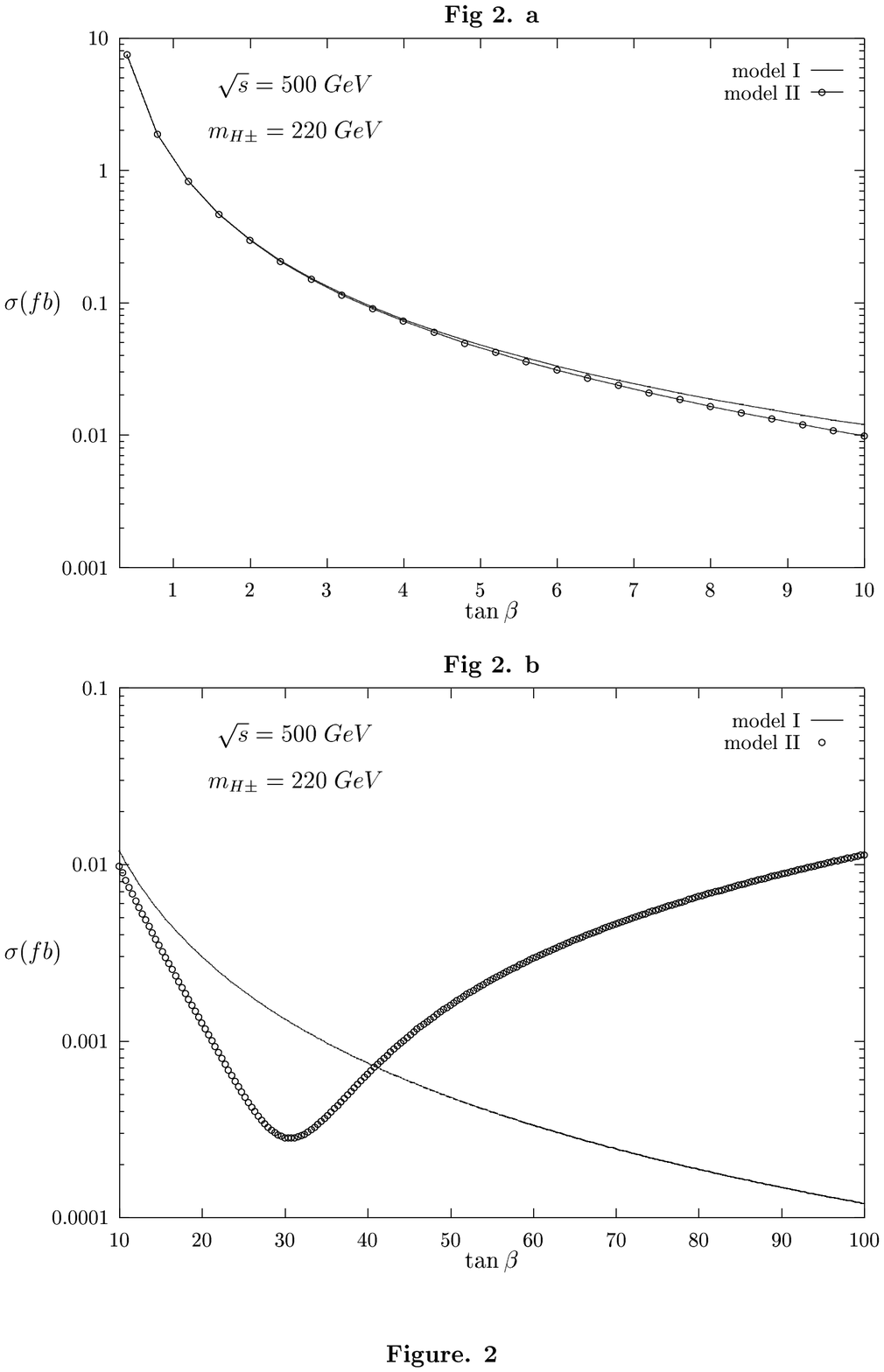}}
\end{picture}
\end{minipage}

\newpage
\begin{minipage}[t]{19.cm}
\setlength{\unitlength}{1.in}
\begin{picture}(1,1)(1.,9.1)
\centerline{\epsffile{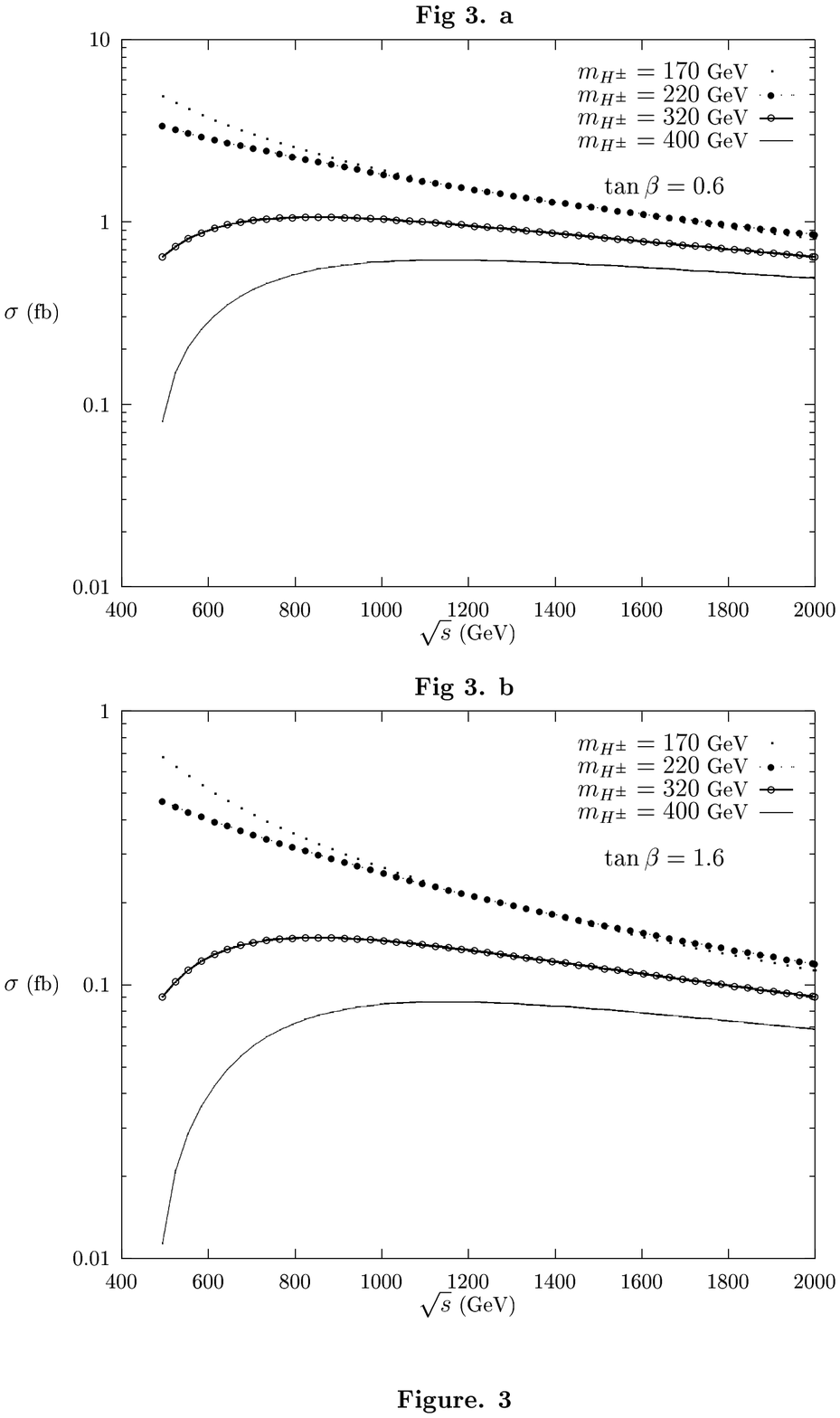}}
\end{picture}
\end{minipage}

\newpage
\begin{minipage}[t]{19.cm}
\setlength{\unitlength}{1.in}
\begin{picture}(1,1)(1.,9.1)
\centerline{\epsffile{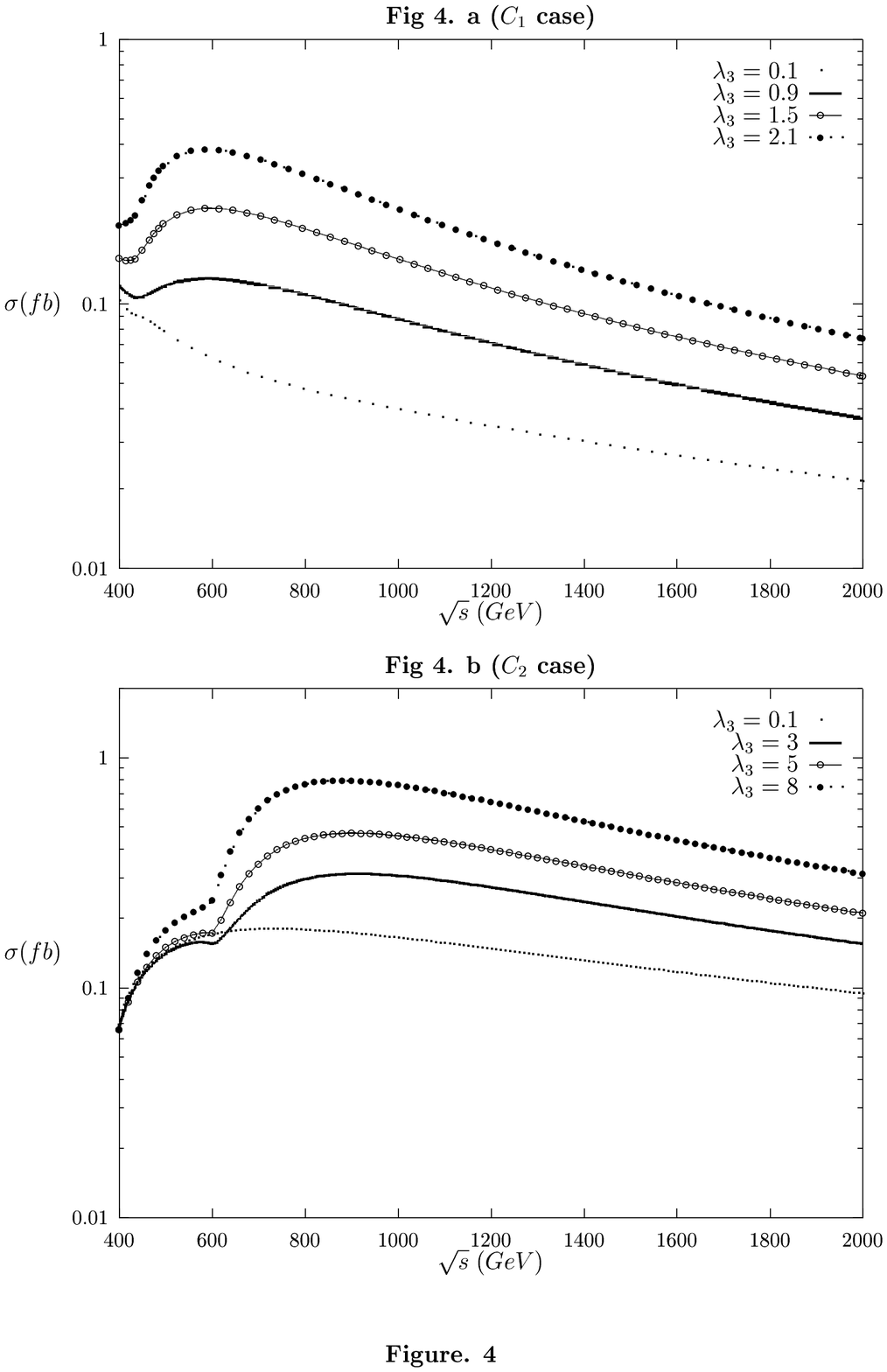}}
\end{picture}
\end{minipage}

\newpage
\begin{minipage}[t]{19.cm}
\setlength{\unitlength}{1.in}
\begin{picture}(1,1)(1.,9.1)
\centerline{\epsffile{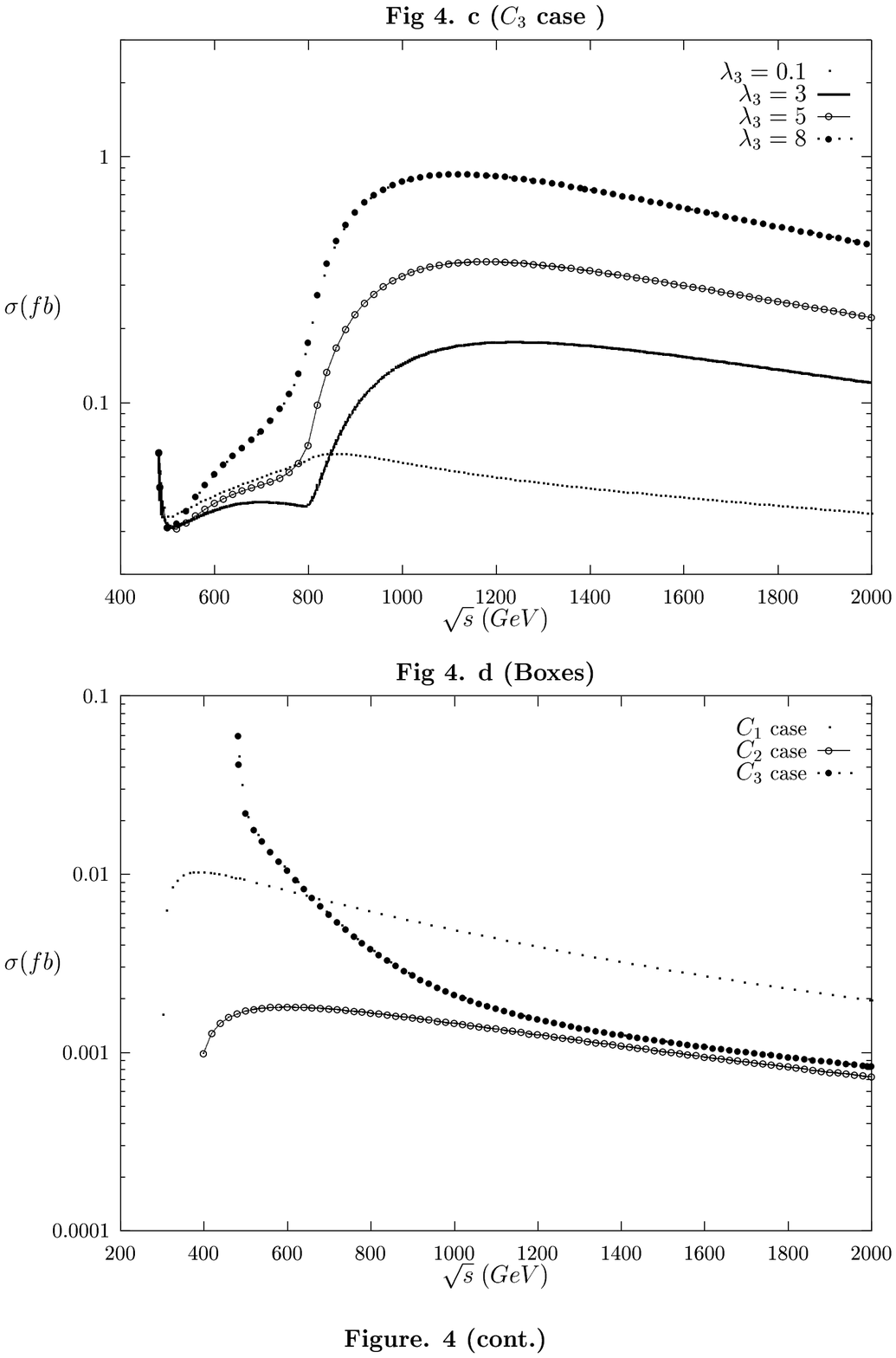}}
\end{picture}
\end{minipage}

\newpage
\begin{minipage}[t]{19.cm}
\setlength{\unitlength}{1.in}
\begin{picture}(1,1)(1.,9.1)
\centerline{\epsffile{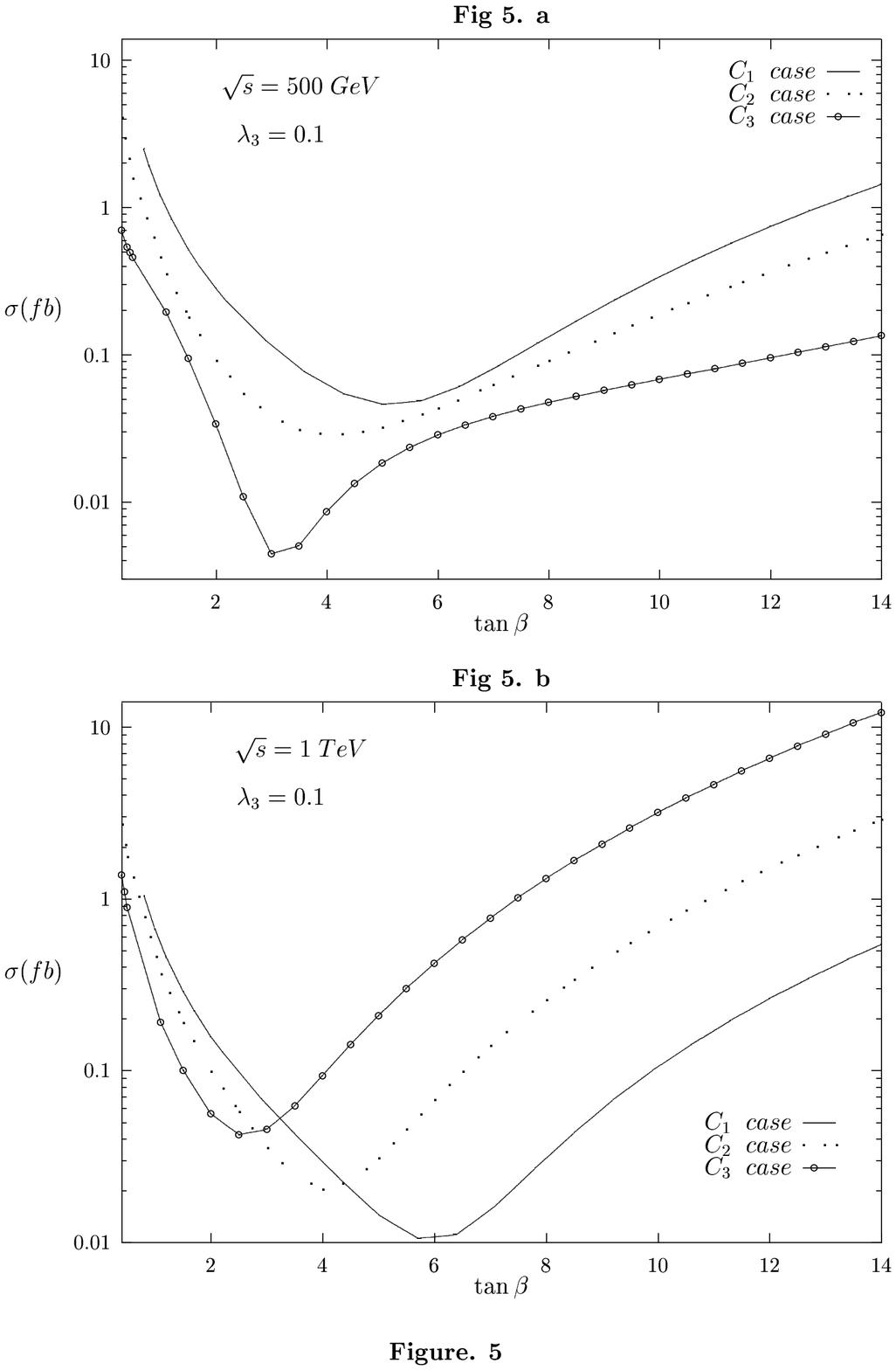}}
\end{picture}
\end{minipage}

\newpage
\begin{minipage}[t]{19.cm}
\setlength{\unitlength}{1.in}
\begin{picture}(1,1)(1.,9.1)
\centerline{\epsffile{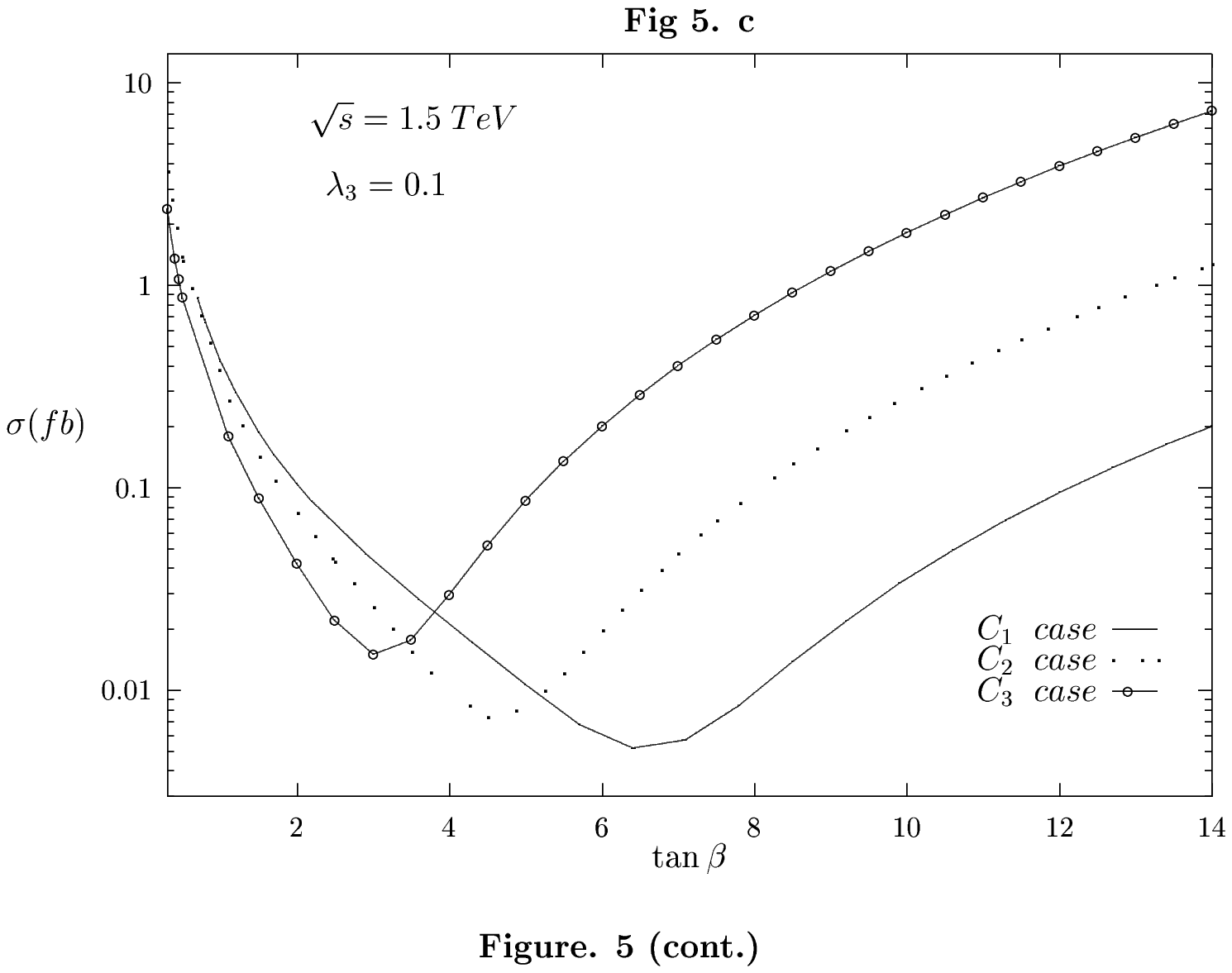}}
\end{picture}
\end{minipage}

\newpage
\begin{minipage}[t]{19.cm}
\setlength{\unitlength}{1.in}
\begin{picture}(1,1)(1.,9.1)
\centerline{\epsffile{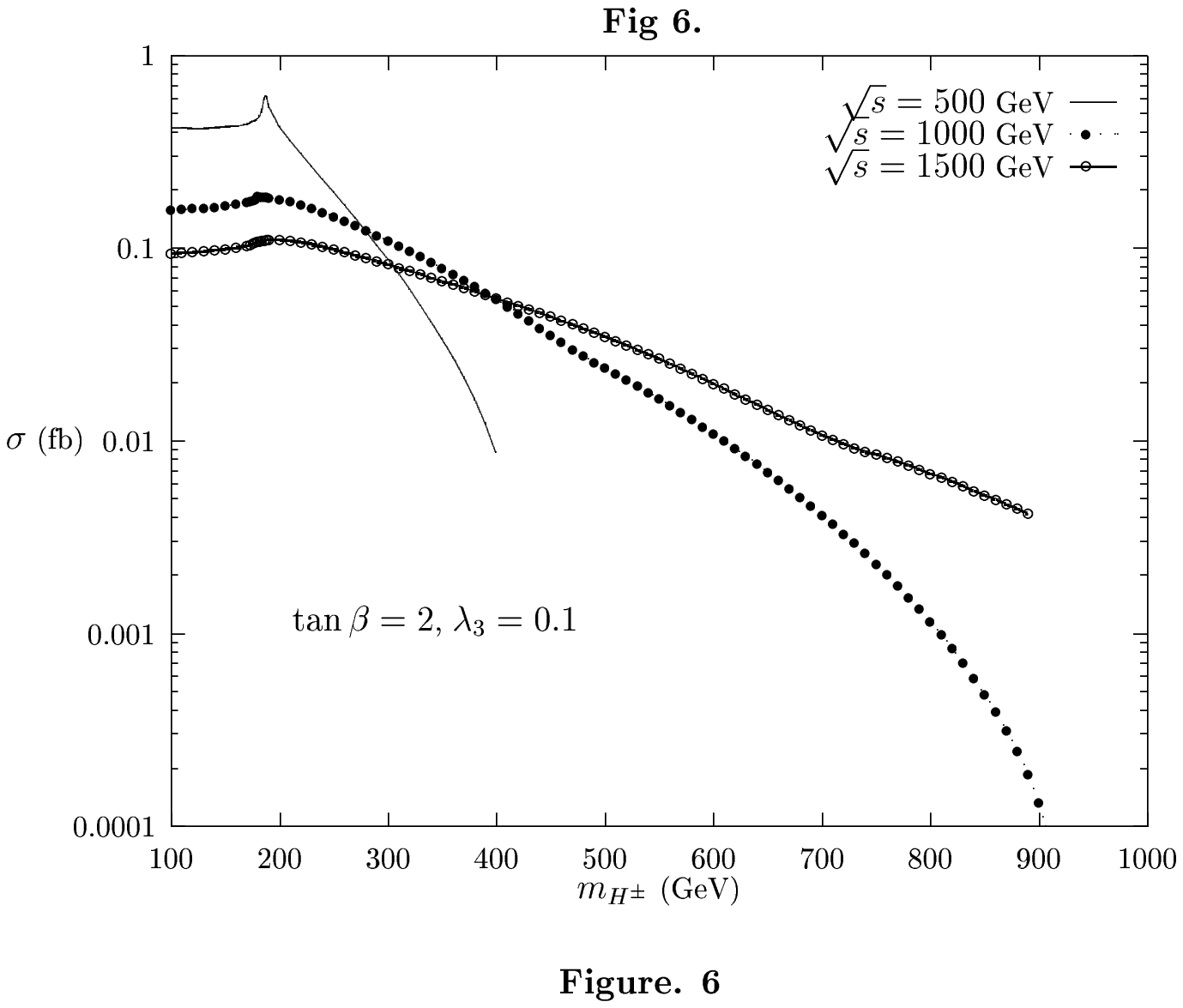}}
\end{picture}
\end{minipage}

\end{document}